\documentclass[12pt]{article}
\usepackage{epsfig}
\usepackage{cite}
\usepackage{float}
\usepackage{mathrsfs}
\usepackage{amsfonts}
\usepackage{array}
\usepackage{amsmath}    % need for subequations
\usepackage{amssymb}   % defines \lesssim, etc
\usepackage{graphicx}   % need for figures
\usepackage{verbatim}   % useful for program listings
\usepackage{color}      % use if color is used in text
\usepackage{subfigure}  % use for side-by-side figures

\newcommand{\mysection}{\setcounter{equation}{0}\section}

\def\beq{\begin{equation}}
\def\eeq{\end{equation}}
\def\beqa{\begin{eqnarray}}
\def\eeqa{\end{eqnarray}}
\def\ba{\begin{eqnarray}}
\def\ea{\end{eqnarray}}

\newlength{\dinwidth} \newlength{\dinmargin}
\begin{document}

\begin{center}
{\Large \bf $tZ'$ production at hadron colliders}
\end{center}
\begin{center}
{\large Marco Guzzi and Nikolaos Kidonakis}\\
\vspace{2mm}
{\it Department of Physics, Kennesaw State University,\\
Kennesaw, GA 30144, USA}
\end{center}

\begin{abstract}
We study the production of a single top quark in association with a heavy extra $Z'$ at hadron colliders in new physics models with and without flavor-changing neutral-current (FCNC) couplings. We use QCD soft-gluon resummation and threshold expansions to calculate higher-order corrections for the total cross section and transverse-momentum distributions for $t Z'$ production. The impact of the uncertainties due to the structure of the proton and scale dependence is also analyzed.
\end{abstract}

\mysection{Introduction}
The top quark is the heaviest particle in the three quark generations.
Its mass of approximately $m_t=172.5$ GeV has been measured with very high accuracy at the Large Hadron Collider (LHC)~\cite{CMS:2017eoz,Kovalchuk:2018ysi,ATLAS:2017lqh}, and being close to that of the Higgs boson it makes the top quark one of the best candidates to probe the Electroweak (EW) sector of the Standard Model (SM) and its extensions.

The accumulated data at the LHC have not yet provided us with evidence of deviations from the SM, but Run II of the LHC and its upgrade to a High Luminosity phase (HL-LHC)~\cite{Azzi:2019yne}, and especially future center-of-mass energy upgrades are going to record a large number of high-energy collision data which will allow us to probe rare processes that may hint at or provide direct evidence of new physics.
In particular, for physics Beyond the Standard Model (BSM) and beyond the LHC, there are several projects going on which provide a synergy of various new-generation
facilities like the Future Circular Collider (FCC)~\cite{FCC} and
the Super proton proton Collider (SppC)~\cite{Tang:2015qga}. With a center-of-mass energy of approximately 100 TeV, these new-generation hadron colliders represent the new frontier for discovery at high energies and will be critical to identify particles with
mass of ${\cal O}(10)$ TeV.
At these energies, we will be able to investigate properties of the Higgs boson and the top quark, and EW symmetry-breaking phenomena with unprecedented precision and sensitivity.
Moreover, the statistics will be enhanced by several orders of magnitude with respect to that of the LHC, and this is going to be ideal to study BSM physics and rare processes.
In this respect, a process of interest is the production of a single top quark in association with a new heavy particle.

Regardless of the type of the new heavy particle, many aspects of this reaction are interesting at quantum-field-theoretical level and because of the phenomenological implications on BSM physics. For example, the kinematics of the final state and decay products can be relevant to investigate extensions of the Higgs sector (two-Higgs-doublet model (2HDM), SUSY, etc.), and of the EW sector with enlarged gauge symmetry.

In this work we shall focus on the production of a top quark in association with an extra $Z'$ vector boson coming from distinct BSM theories, and we will analyze higher-order QCD corrections to this process due to soft gluon emissions.

Extra vector gauge bosons, generically referred to as extra $Z'$s, are almost ubiquitous in extensions of the EW sector of the SM.
$Z'$s are associated with additional abelian $U'(1)$ gauge symmetries which
were suggested in SM extensions such as left-right symmetric models, Grand Unified Theories (GUTs) and string-inspired constructions (see Refs.~\cite{Langacker:2008yv,Rizzo:2006nw,Leike:1998wr,Hewett:1988xc,Komachenko:1989qn,Babu:1997st} for reviews and references).
In the past decade, $Z'$ gauge bosons at the TeV scale gathered considerable attention in theoretical calculations (including parton-shower) ~\cite{Adelman:2012py,Fuks:2007gk,Jezo:2014wra,Bonciani:2015hgv,Fuks:2017vtl,Araz:2017wbp} and triggered a vigorous program of experimental searches at the LHC.
At high energies, $Z'$s can in principle have different signatures: they can be produced as intermediate resonances in Drell-Yan processes as well as in association with another SM vector or scalar boson, or in association with a jet or single top quark such as in the case of $pp\rightarrow t Z'$.

The dynamics of this process is non-trivial because of several hard scales entering the cross section.
In fact, in high-energy reactions in which the final-state heavy particle has a mass much heavier than the top quark mass,
$m_t$, the cross section is affected by large (collinear) logarithmic contributions of the
type $\alpha_s^n\log^n {\left(Q^2/m_t^2\right)}$ (where $Q\approx m_{Z'}$, the $Z'$-boson mass,
and $\alpha_s$ is the QCD coupling constant) that can spoil the convergence of the perturbative
series in calculations at fixed order \cite{Dicus:1998hs}. Therefore, there is the necessity of
resumming these logarithmic contributions using DGLAP evolution defining a top-quark parton distribution function (PDF) inside the proton.
When a higher energy scale $Q \approx m_{Z'}$ involving a heavy final state is such that $m_{Z'}\gg m_t$,
the top quark can be considered essentially massless and an active flavor inside the proton.
Details of factorization schemes with different number of flavors with consistent treatment of the
top quark as a massless degree of freedom at high energies are discussed in Refs.~\cite{Dawson:2014pea,Han:2014nja} and references therein.
In particular, QCD factorization with initial-state heavy flavors is discussed in Refs.~\cite{Aivazis:1993pi,Collins:1998rz,Kramer:2000hn,Tung:2001mv,Guzzi:2011ew}.

In processes with very heavy final states the near-threshold kinematic region becomes particularly important. Soft-gluon corrections typically become large and dominant in such circumstances. Therefore, the $K$-factors can become quite large and it is important to include these corrections in making theoretical predictions. In this study, we adopt and extend the soft-gluon resummation formalism used in \cite{Kidonakis:2003sc,Kidonakis:2017mfy,Forslund:2018qcp} for $tZ$ and $t\gamma$ production (see also applications to top-antitop pair production~\cite{NKttbar} and single-top production \cite{NKsingletop}, and a review in \cite{Kidonakis:2018ybz}) to calculate approximate next-to-next-to-leading order (aNNLO) cross sections for $tZ'$ associated production in two case scenarios: {\it i}) the case of $Z'$s with flavor-changing anomalous couplings, {\it ii}) the case of $Z'$s originating from low-energy realizations of string models.
We explore the impact of the corrections due to multiple emission of soft-gluons as well as the cross section suppression due to $Z'$s of different mass and couplings.
Moreover, we analyze the uncertainties in the cross section associated to the PDFs of the initial state protons and to the factorization $\mu_F$ and renormalization $\mu_R$ scales.
Finally, we generate prospects for the cross section for the case studies mentioned above, at future generation ultra-high energy colliders.

The paper is organized as follows. In Sec.~\ref{EffectiveLagrangianFCNC} we discuss the BSM effective Lagrangians, couplings, and leading-order cross sections. In Sec.~\ref{Soft-gluons} we illustrate the soft-gluon formalism and calculate the higher-order corrections. In
Sec.~\ref{Pheno} we present results for the total cross sections and top-quark transverse-momentum ($p_T$) distributions in $tZ'$ production via the processes $gu\rightarrow tZ'$ and $gc \rightarrow tZ'$ with anomalous couplings, and also via the process $gt \rightarrow tZ'$. We conclude in Sec.~\ref{Conclusions}.

\mysection{\label{EffectiveLagrangianFCNC} Effective Lagrangians}

\subsection{Lagrangian for FCNC $Z'$s}

\begin{figure}[th!]
\begin{center}
\includegraphics[width=10cm]{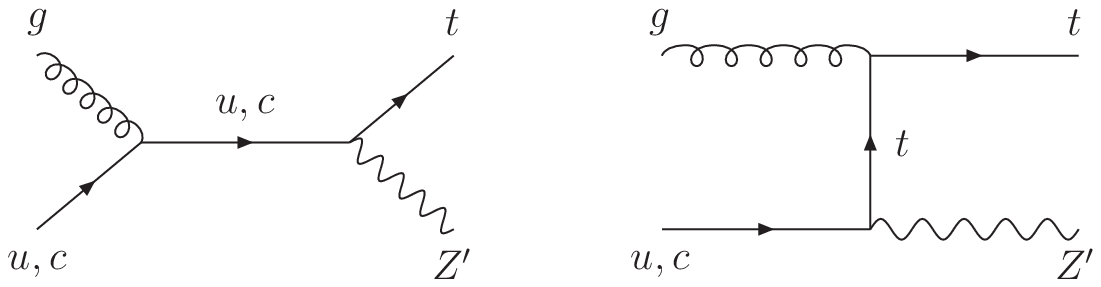}
\caption{Leading-order diagrams for $gu\rightarrow tZ'$ with anomalous $t$-$u$-$Z'$ coupling and $gc\rightarrow tZ'$ with anomalous $t$-$c$-$Z'$ coupling.}
\label{gqtZp}
\end{center}
\end{figure}

An FCNC term in the Lagrangian that includes the anomalous coupling of a $t,q$ pair to a $Z'$ boson is given by
\begin{equation}
{\cal L}_{FCNC} =    \frac{1}{ \Lambda } \,
\kappa_{tqZ'} \, e \, \bar t \, \sigma_{\mu\nu} \, q \, F^{\mu\nu}_{Z'} + h.c.,
\label{Langrangian}
\end{equation}
where $\kappa_{tqZ'}$ is the anomalous $t$-$q$-$Z'$ coupling, with
$q$ an up or charm quark; $e$ is the electron charge;
$\Lambda$ is an effective new physics scale in the few TeV's range;
$F^{\mu\nu}_{Z'}$  is the $Z'$ field tensor; and
$\sigma_{\mu \nu}=(i/2)(\gamma_{\mu}\gamma_{\nu}
-\gamma_{\nu}\gamma_{\mu})$ with $\gamma_{\mu}$ the Dirac matrices.

The partonic processes involved are $gu \rightarrow tZ'$ and $gc \rightarrow tZ'$.
Leading-order diagrams for these processes are shown in Fig. \ref{gqtZp}.
Related processes involving $Z$ bosons with anomalous couplings were studied in Refs. \cite{Kidonakis:2003sc,Kidonakis:2017mfy}.

\subsection{\label{stringyZprime} Lagrangian for string-inspired $Z'$s }

The Lagrangian for a $Z'$ coming from string-inspired models is given below, where
we adopt the notation introduced in Refs.~\cite{Coriano:2008wf,Faraggi:2015iaa}. Here we report the most basic definitions for completeness.

The fermion-fermion-$Z^{\prime}$ interaction is given by
\ba
\sum_{i=L,R} z_{t,i} g_{Z'} \bar{t}_i \gamma^{\mu} t_i Z_{\mu}^{\prime},
\ea
where the coefficients $z_{t,L}$, and $z_{t,R}$ are the charges of the left- and right-handed top quarks respectively.
The $Z'$ coupling is indicated by $g_{Z'}$.

The mass of the $Z$ gauge boson is parametrized in terms of the
vacuum expectation values (vev's) of the Higgs sector $v_{H_1}$, $v_{H_2}$ as follows
\ba
&&m_Z^2=\frac{g^2}{4
\cos^2\theta_W}(v_{H_1}^2+v_{H_2}^2)\left[1+O(\varepsilon^2)\right] \, ,
\nonumber\\
&&\varepsilon=\frac{\delta m^2_{Z
Z^{\prime}}}{m^2_{Z^{\prime}}-m^2_{Z}} \, ,
\nonumber\\
&&\delta m^2_{Z Z^{\prime}}=-\frac{g g_{Z'}}{4\cos\theta_W}(z_{H_1}^2
v_{H_1}^2+z_{H_2}^2v_{H_2}^2) \, ,
\ea
where the mixing parameter $\varepsilon$ is defined perturbatively, $z_{H_1}$ and $z_{H_2}$ are the charges of the Higges,
$g=e/\sin\theta_W$, $g_Y= e/\cos\theta_W$, and $\theta_W$ is the Weinberg angle. We consider $m_{Z'}$ as a free parameter in the TeV's range.
We restrict our attention to the interaction Lagrangian for the top-quark sector only, which is written as
\ba
&&{\mathcal{L}}_{int}=
\bar{t}_{L} N^{Z^{\prime}}_{L}\gamma^{\mu}t_{L} Z^{\prime}_{\mu}
+\bar{t}_{R}N^{Z^{\prime}}_{R}\gamma^{\mu}t_{R} Z^{\prime}_{\mu}\,,
\ea
where the left-handed (L) and right-handed (R) couplings are
\ba
&&N^{Z^{\prime}}_{L}=-i\left(-g\cos\theta_W T_{3,L}\varepsilon
+g_Y\sin\theta_W
\frac{ Y_{t,L}}{2}\varepsilon+g_{Z'}\frac{ z_{t,L}}{2}\right) \, ,
\nonumber\\
&&N^{Z'}_{R}=-i\left(
g_Y \sin\theta_W \frac{ Y_{t,R}}{2}\varepsilon +g_{Z'}
\frac{z_{t,R}}{2}\right) \,,
\ea
where $Y_{t,L/R}$ is the hypercharge and $T_{3,L}$ is the weak isospin.

\begin{figure}[th!]
\begin{center}
\includegraphics[width=10cm]{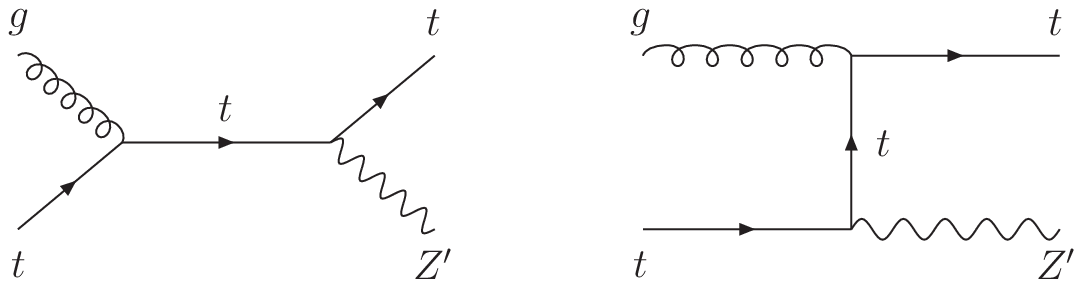}
\caption{Leading-order diagrams for $gt\rightarrow tZ'$.}
\label{gttZp}
\end{center}
\end{figure}

Based on this Lagrangian, we will study below the process $gt \rightarrow tZ'$. The leading-order diagrams for this process are shown in Fig. \ref{gttZp}.

\subsection{Hadronic cross section}

The hadronic cross section for $p(P_1)+p(P_2)\rightarrow t(p_t)+Z'(p_{Z'})$ is expressed in terms of Mandelstam variables
\ba
S=(P_1+P_2)^2\,, ~~T=(P_1-p_t)^2  \,, ~~U=(P_2-p_t)^2 \,, ~~S_4 = S + T+ U-m_t^2-m_{Z'}^2\,.
\ea
We also define $T_1=T  -m_t^2$ and $U_1=U -m_t^2$.

The factorized differential cross section can be written as
\ba
S^2\frac{d^2 \sigma(S,T_1,U_1)}{dT_1 ~dU_1} &=& \sum_{i,j=q,g} \int_{x_1^-}^{1} \frac{dx_1}{x_1}
\int_{x_2^-}^{1}\frac{dx_2}{x_2} f_{i/p_1}(x_1,\mu_F^2) f_{j/p_2}(x_2,\mu_F^2)
\nonumber\\
&&\times \, \hat{\sigma}_{ij\rightarrow tZ'}(s,t_1,u_1,m_t^2,m^2_{Z'},\mu_F^2,\alpha_s(\mu_R^2)) + {\cal O}(\Lambda_{QCD}^2/\Lambda^2)
\label{diffXsec}
\ea
where $f_{j/p}(x,\mu_F^2)$ is the parton distribution function representing the probability of finding the parton $j$ in proton $p$, $\mu_F$ and $\mu_R$ are the factorization and renormalization scales respectively, and $\hat\sigma_{ij\rightarrow tZ'}$ is the hard scattering cross section.
Here, $\Lambda_{QCD}$ is the QCD scale while the scale $\Lambda$ is of the order of $m_{Z'}$, and power suppressed terms $\Lambda_{QCD}^2/\Lambda^2$ are neglected.
In our numerical results in Sec. 4 we set $\mu_F=\mu_R=\mu$.

The lower integration limits in the factorization formula are given by
\beq
x_1^-= -\frac{U_1}{S + T_1} \,, ~~ x_2^-= -\frac{x_1 T_1}{x_1 S + U_1}.
\eeq
The double-differential cross section in Eq. (\ref{diffXsec}) can be written
in terms of the transverse momentum $p_T$ of the top quark and its rapidity $y$ using
\beq
T_1= - \sqrt{S} ~m_T e^{-y}\,,~~ U_1= - \sqrt{S} ~m_T e^y\,,
\eeq
where the transverse mass $m_T$ is defined as $m_T=\sqrt{p_T^2 + m_t^2}$.

\subsection{Leading-order cross sections}

For the partonic process $g(p_g)+q(p_q) \rightarrow t(p_t)+Z'(p_{Z'})$,
we define the kinematical variables $s=(p_g+p_q)^2$,
$t=(p_g-p_t)^2$, and $u=(p_q-p_t)^2$.

The leading-order (LO) double-differential partonic cross section for $g q \rightarrow tZ'$, with $q$ and up or charm quark, via anomalous couplings is
\beq
\frac{d^2{\hat\sigma^{(0)}}_{gq \rightarrow t Z'}}{dt \, du}
=F^{\rm LO}_{gq \rightarrow t Z'} \, \delta(s_4) \, ,
\label{LO}
\eeq
where
\beqa
F^{\rm LO}_{gq \rightarrow t Z'}&=&
\frac{2 \pi \alpha \alpha_s \kappa_{tqZ'}^2}{3s^3 (t-m_t^2)^2 \Lambda^2}
\left\{2 m_t^8 -m_t^6(3 m_{Z'}^2+4 s+2 t) \right.
\nonumber \\ &&
{}+m_t^4\left[2m_{Z'}^4-m_{Z'}^2(2s+t)+2(s^2+4st+t^2)\right]
\nonumber \\ &&
{}+m_t^2\left[2m_{Z'}^6-4m_{Z'}^4t+m_{Z'}^2(s+t)(s+5t)
-2t(3s^2+6st+t^2)\right]
\nonumber \\ && \left.
{}-t\left[2 m_{Z'}^6-2m_{Z'}^4(s+t)
+m_{Z'}^2(s+t)^2-4 s t(s+t)\right]\right\} \, ,
\eeqa
with $\alpha=e^2/(4\pi)$.

For the partonic process $g(p_g)+t(p_q) \rightarrow t(p_t)+Z'(p_{Z'})$,
we again define the kinematical variables $s=(p_g+p_q)^2$,
$t=(p_g-p_t)^2$, and $u=(p_q-p_t)^2$.
The LO cross section for $g t \rightarrow t Z'$ is given by
\begin{eqnarray}
&&F^{\rm LO}_{g t\rightarrow t Z'}=
4 \frac{4\pi \alpha_s}{m_{Z'}^2 N_c (s-m_t^2)^2 (t-m_t^2)^2}
\left\{g_{AtZ'}^2 \left[2 m_{Z'}^4 ((2m_t^2 s t-5 m_t^4 (s+t)+6 m_t^6)+s t(s+t))
\right.\right.
\nonumber\\
&&\left.\left.
   +2 m_{Z'}^6 (s-m_t^2)(m_t^2-t)-m_{Z'}^2 (m_t^4 (s-3 t) (3 s-t)
   -12 m_t^6 (s+t)-m_t^2 (s+t) (-6 st+s^2+t^2)
\right.\right.
\nonumber\\
&&\left.\left.
   +18 m_t^8+s t (s^2+t^2))-2 m_t^2 (s-m_t^2) (t-m_t^2) (-2 m_t^2+s+t)^2\right]
\right.
\nonumber\\
&&\left.
   +g_{VtZ'}^2 m_{Z'}^2 \left[-s t
   (2 m_{Z'}^4+s^2+t^2-2 m_{Z'}^2
   (s+t))
  -m_t^4 (-2 m_{Z'}^2 (s+t)+2 m_{Z'}^4+14 st+3 s^2+3 t^2)
\right.\right.
\nonumber\\
&&\left.\left.
   +m_t^2 (-8 m_{Z'}^2 s t+2
   m_{Z'}^4 (s+t)+(s+t) (6 s t+s^2+t^2))+6
   m_t^8\right]\right\} \,,
   \nonumber\\
\end{eqnarray}

where the vector and axial coupling of the $Z^{\prime}$ boson to the top quark are

\begin{eqnarray}
&&\frac{-i g}{4 c_w}\gamma^{\mu} g_{VtZ^{\prime}}=\frac{-i g}{c_w}
\frac{1}{2}\left[ -\varepsilon c_w^2 T_3^{L}
+\varepsilon s_w^2(\frac{Y_{t,L}}{2}+\frac{Y_{t,R}}{2})
+\frac{g_{Z'}}{g}c_w(\frac{z_{t,L}}{2}+
\frac{z_{t,R}}{2})\right]\gamma^{\mu}
\nonumber\\
&&\frac{-i g}{4 c_w}\gamma^{\mu}\gamma^{5} g_{AtZ^{\prime}}=\frac{-i
g}{c_w} \frac{1}{2}\left[ \varepsilon c_w^2 T_3^{L}
+\varepsilon s_w^2(\frac{Y_{t,R}}{2}-\frac{Y_{t,L}}{2})
+\frac{g_{Z'}}{g}c_w(\frac{z_{t,R}}{2}-
\frac{z_{t,L}}{2})\right]\gamma^{\mu}\gamma^{5},
\nonumber\\
\end{eqnarray}
where we set $\sin\theta_W=s_w$ and $\cos\theta_W=c_w$ for brevity.

\mysection{\label{Soft-gluons} Soft-gluon corrections}

We next describe the formalism and procedure for calculating soft-gluon corrections in the cross section for $tZ'$ production.
For the processes $gq \rightarrow tZ'$ and $gt \rightarrow tZ'$, we defined the usual
kinematical variables $s$, $t$, and $u$, in the previous section. We can also define  a threshold kinematical variable, $s_4=s+t+u-m_t^2-m_{Z'}^2$, that measures distance from partonic threshold, and vanishes at partonic threshold where there is no energy available for additional radiation. More specifically, $s_4$ is the squared invariant mass of additional final-state radiation.
We also define $t_1=t-m_t^2$, $t_2=t-m_{Z'}^2$, $u_1=u-m_t^2$, and $u_2=u-m_{Z'}^2$.

The resummation of soft-gluon contributions to the partonic process follows from the factorization of the cross section as a product of functions that describe soft and collinear emission. Taking the Laplace transform
${\hat \sigma}(N)=\int (ds_4/s) \;  e^{-N s_4/s} {\hat \sigma}(s_4)$,
we have a factorized expression in $4-\epsilon$ dimensions,
\beq
\frac{d^2{\hat \sigma}_{gq \rightarrow tZ'}(N,\epsilon)}{dt \, du}=
H_{gq \rightarrow tZ'} \left(\alpha_s(\mu)\right)\; S_{gq \rightarrow tZ'}
\left(\frac{m_t}{N \mu},\alpha_s(\mu) \right)\;
\prod_{i=g,q} J_i\left (N,\mu,\epsilon \right)
\label{factsigma}
\eeq
where $H_{gq \rightarrow tZ'}$ is a hard function,
$S_{gq \rightarrow tZ'}$ is a soft function for
noncollinear soft-gluon emission,
and $J_i$ are jet functions for
soft and collinear emission from the incoming quark and gluon.
Our considerations are identical for all three processes to be studied in this paper, i.e. $gu\rightarrow tZ'$, $gc \rightarrow tZ'$, and $gt \rightarrow tZ'$.

The dependence of the soft function $S_{gq \rightarrow tZ}$ on $N$
is resummed via renormalization
group evolution \cite{Kidonakis:2017mfy,Forslund:2018qcp,NKttbar,NKsingletop,Kidonakis:1997gm},
\beq
S^b_{gq \rightarrow tZ'}=(Z^S)^* \; S_{gq \rightarrow tZ'} \, Z^S \, ,
\eeq
with $S^b_{gq \rightarrow tZ}$ the unrenormalized quantity and $Z^S$ a renormalization constant. The function $S_{gq \rightarrow tZ}$ obeys the renormalization group equation
\beq
\left(\mu \frac{\partial}{\partial \mu}
+\beta(g_s, \epsilon)\frac{\partial}{\partial g_s}\right)\,S_{gq \rightarrow tZ'}
=-2 \, S_{gq \rightarrow tZ'} \, \Gamma^S_{gq \rightarrow tZ'} \, ,
\eeq
where $g_s^2=4\pi\alpha_s$,
$\beta(g_s, \epsilon)=-g_s \epsilon/2 + \beta(g_s)$
with $\beta(g_s)$ the QCD beta function, and
\beq
\Gamma^S_{gq \rightarrow tZ'}=\frac{dZ^S}{d\ln\mu} (Z^S)^{-1}
=\beta(g_s, \epsilon)  \frac{\partial Z^S}{\partial g_s} (Z^S)^{-1}
\eeq
is the soft anomalous dimension that determines the evolution of $S_{gq \rightarrow tZ}$. The soft anomalous dimension $\Gamma^S_{gq \rightarrow tZ}$ is calculated in dimensional regularization from the coefficients of the ultraviolet poles of the loop diagrams involved in the process \cite{Kidonakis:2017mfy,Forslund:2018qcp,NKttbar,NKsingletop,Kidonakis:2018ybz,Kidonakis:1997gm,Kidonakis:2009ev,Kidonakis:2019nqa}.

The resummed partonic cross section in moment space
is then given by
\beqa
\frac{d^2{\hat{\sigma}}^{\rm resum}_{gq \rightarrow tZ}(N)}{dt \, du} &=&
\exp\left[\sum_{i=g,q} E_i(N_i)\right]
H_{gq \rightarrow tZ'}
\left(\alpha_s(\sqrt{s})\right) \;
S_{gq \rightarrow tZ'}\left(\alpha_s(\sqrt{s}/{\tilde N'})
\right)
\nonumber \\ &&
\times \exp \left[2\int_{\sqrt{s}}^{{\sqrt{s}}/{\tilde N'}}
\frac{d\mu}{\mu}\; \Gamma^S_{gq \rightarrow tZ'}
\left(\alpha_s(\mu)\right)\right]  \, .
\label{resum}
\eeqa
Soft-gluon resummation is the exponentiation of logarithms of $N$. The first exponent in Eq. (\ref{resum}) includes soft and collinear
corrections \cite{Sterman:1986aj, Catani:1989ne} from the incoming partons,
and can be found explicitly in \cite{NKsingletop}.

We write the perturbative series for the soft anomalous dimension for
$gq \rightarrow tZ'$ as
$\Gamma^S_{gq \rightarrow tZ'}=\sum_{n=1}^{\infty}(\alpha_s/\pi)^n
\Gamma^{S \, (n)}_{gq \rightarrow tZ'}$.
To achieve resummation at next-to-leading-logarithm (NLL) accuracy we require
the one-loop result which is given, in Feynman gauge, by
\beq
\Gamma^{S\, (1)}_{gq \rightarrow tZ}=
C_F \left[\ln\left(\frac{-u_1}{m_t\sqrt{s}}\right)
-\frac{1}{2}\right] +\frac{C_A}{2} \ln\left(\frac{t_1}{u_1}\right) \, ,
\label{tZ1l}
\eeq
with color factors $C_F=(N_c^2-1)/(2N_c)$ and $C_A=N_c$,
where $N_c=3$ is the number of colors.

Upon expanding the resummed cross section to fixed order and inverting from the transform moment space back to momentum space, the logarithms of $N$ produce ``plus'' distributions of logarithms of $s_4/m_{Z'}^2$. The highest power of these logarithms is 1 at NLO and 3 at NNLO.

The NLO soft-gluon corrections for $g q \rightarrow tZ'$ are
\beqa
\frac{d^2{\hat\sigma}^{(1)}_{gq\rightarrow t Z'}}{dt \, du}
&=&F^{\rm LO}_{gq \rightarrow t Z'}
\frac{\alpha_s(\mu_R^2)}{\pi} \left\{
2 (C_F+C_A) \left[\frac{\ln(s_4/m_{Z'}^2)}{s_4}\right]_+ \right.
\nonumber \\ && \hspace{-23mm}
{}+\left[2 C_F \ln\left(\frac{u_1}{t_2}\right)
+C_F \ln\left(\frac{m_{Z'}^2}{m_t^2}\right)-C_F
+C_A \ln\left(\frac{t_1}{u_1}\right)
+C_A \ln\left(\frac{s m_{Z'}^2}{u_2^2}\right) \right.
\nonumber \\ && \hspace{-15mm} \left.
{}-(C_F+C_A)\ln\left(\frac{\mu_F^2}{m_{Z'}^2}\right)\right]
\left[\frac{1}{s_4}\right]_+
\nonumber \\ && \hspace{-23mm} \left.
{}+\left[\left(C_F \ln\left(\frac{-t_2}{m_{Z'}^2}\right)
+C_A \ln\left(\frac{-u_2}{m_{Z'}^2}\right)
-\frac{3}{4}C_F\right)\ln\left(\frac{\mu_F^2}{m_{Z'}^2}\right)
-\frac{\beta_0}{4}\ln\left(\frac{\mu_F^2}{\mu_R^2}\right)\right]
\delta(s_4)\right\} \, ,
\label{NLOgqtZp}
\eeqa
where $\beta_0=(11C_A-2n_f)/3$ is the lowest-order QCD $\beta$ function, with $n_f$ the number of light quark flavors.
We set $n_f=5$ for $gu\rightarrow tZ'$ and $gc\rightarrow tZ'$, and $n_f=6$ for $gt\rightarrow tZ'$.
The leading logarithms in the NLO expansion are the $[\ln(s_4/m_{Z'}^2)/s_4]_+$ terms while the NLL are the $[1/s_4]_+$ terms.
In addition, at NLL we determine in Eq. (\ref{NLOgqtZp}) the $\delta(s_4)$ terms involving the scale.
In top-quark production processes, the NLO soft-gluon corrections approximate very well the complete NLO
corrections \cite{Kidonakis:2017mfy,Forslund:2018qcp,NKttbar,NKsingletop,Kidonakis:2018ybz}.
We denote the sum of the LO cross section and the NLO soft-gluon corrections as approximate NLO (aNLO).

The NNLO soft-gluon corrections for $g q \rightarrow tZ'$ are
\beqa
\frac{d^2{\hat\sigma}^{(2)}_{gq\rightarrow t Z'}}{dt \, du}
&=&F^{\rm LO}_{gq \rightarrow t Z'}
\frac{\alpha_s^2(\mu_R^2)}{\pi^2} \left\{
2(C_F+C_A)^2 \left[\frac{\ln^3(s_4/m_{Z'}^2)}{s_4}\right]_+ \right.
\nonumber \\ && \hspace{-25mm}
{}+3(C_F+C_A)\left[2 C_F \ln\left(\frac{u_1}{t_2}\right)
+C_F \ln\left(\frac{m_{Z'}^2}{m_t^2}\right)-C_F
+C_A \ln\left(\frac{t_1}{u_1}\right)
+C_A \ln\left(\frac{s m_{Z'}^2}{u_2^2}\right) \right.
\nonumber \\ && \left.
{}-(C_F+C_A)\ln\left(\frac{\mu_F^2}{m_{Z'}^2}\right) -\frac{\beta_0}{6}\right]
\left[\frac{\ln^2(s_4/m_{Z'}^2)}{s_4}\right]_+
\nonumber \\ && \hspace{-25mm}
{}+2(C_F+C_A)\left[\left(3C_F \ln\left(\frac{-t_2}{m_{Z'}^2}\right)
-2C_F \ln\left(\frac{-u_1}{m_{Z'}^2}\right)
-C_F \ln\left(\frac{m_{Z'}^2}{m_t^2}\right)+\frac{C_F}{4}
+3C_A \ln\left(\frac{-u_2}{m_{Z'}^2}\right) \right. \right.
\nonumber \\ &&  \left.
{}+C_A \ln\left(\frac{u_1 m_{Z'}^2}{t_1 s}\right)-\frac{\beta_0}{4}\right)
\ln\left(\frac{\mu_F^2}{m_{Z'}^2}\right)
+\frac{\beta_0}{2} \ln\left(\frac{\mu_R^2}{m_{Z'}^2}\right)
+\frac{1}{2}(C_F+C_A) \ln^2\left(\frac{\mu_F^2}{m_{Z'}^2}\right)
\nonumber \\ &&  \left.
-2(C_F+C_A) \zeta_2 \right] \left[\frac{\ln(s_4/m_{Z'}^2)}{s_4}\right]_+
\nonumber \\ && \hspace{-25mm}
{}+(C_F+C_A)\left[\left(\frac{3\beta_0}{8}+\frac{3}{4}C_F
-C_F\ln\left(\frac{-t_2}{m_{Z'}^2}\right)
-C_A\ln\left(\frac{-u_2}{m_{Z'}^2}\right)\right)\ln^2\left(\frac{\mu_F^2}{m_{Z'}^2}
\right) -\frac{\beta_0}{2}\ln\left(\frac{\mu_F^2}{m_{Z'}^2}\right)\ln\left(\frac{\mu_R^2}{m_{Z'}^2}\right)\right.
\nonumber \\ &&
{}-2 \zeta_2 \left(2 C_F \ln\left(\frac{u_1}{t_2}\right)
+C_F \ln\left(\frac{m_{Z'}^2}{m_t^2}\right)-C_F
+C_A \ln\left(\frac{t_1}{u_1}\right)+C_A \ln\left(\frac{s m_{Z'}^2}{u_2^2}\right)
\right.
\nonumber \\ &&  \left. \left. \left.
{}-(C_F+C_A)\ln\left(\frac{\mu_F^2}{m_{Z'}^2}\right)\right)
+4(C_F+C_A)\zeta_3 \right]
\left[\frac{1}{s_4}\right]_+ \right\} \, .
\nonumber \\
\label{NNLOgqtZp}
\eeqa
The leading logarithms in the NNLO expansion are the $[\ln^3(s_4/m_{Z'}^2)/s_4]_+$ terms while the NLL are the
$[\ln^2(s_4/m_{Z'}^2)/s_4]_+$ terms. Moreover, at NLL we determine in Eq. (\ref{NNLOgqtZp}) additional terms involving
the scale. The cross section with the inclusion of the soft-gluon corrections through NNLO is denoted as approximate NNLO (aNNLO).

\mysection{\label{Pheno} Phenomenological analysis}

In the following sections we present the results of our phenomenological analysis in which we
investigate the impact of the QCD corrections due to soft gluon emissions to the production of
a single top quark in association with a $Z'$ for the case studies previously discussed.

According to recent LHC Run II exclusion limits~\cite{Sirunyan:2018exx,Aad:2019fac}, extra neutral currents with masses $m_{Z'}\lesssim$ 4 TeV are disfavoured.
In our analysis we consider final-state $Z'$s with masses ranging from 1 to 8 TeV where lighter $Z'$ masses are still included, because we wish to illustrate
the behavior of the cross section and its scaling with the different phase-space suppression due to a final state with $Z'$ masses from low to high.

\subsection{Comparison with existing results at NLO}

We first illustrate a comparison of our aNLO calculation against other existing results at NLO.
Then we discuss the matching of our aNNLO calculation to the exact NLO at fixed order in QCD.
To validate the formalism at aNLO, we use $tZ$ production at the LHC in the presence of FCNC and compare the total cross
section and scale dependence for the $gu \rightarrow tZ$ channel at NLO to the results of Ref.~\cite{Li:2011ek}.
The comparison is summarized in Table~\ref{Comparison-Table} and was already documented in Ref.~\cite{Kidonakis:2017mfy}.
\begin{table}[]
\begin{center}
\begin{tabular}{l l l l l}
\hline
\hline
 $\sigma_0^\textrm{aNLO}$ &  $\sigma_0^\textrm{NLO}$ (Ref.~\cite{Li:2011ek}) & $(\sigma(\mu)/\sigma_0)_{ \textrm{aNLO}}$ & $(\sigma(\mu)/\sigma_0)_{ \textrm{NLO}}$ (Ref.~\cite{Li:2011ek})&  \\
\hline
  22.55 pb & 22.5 pb &  ${}^{0.912 ~~ \mu=2 (m_Z + m_t)}_{1.103 ~~ \mu = (m_Z + m_t)/2}$ & ${}^{0.913~~ \mu=2 (m_Z + m_t)}_{1.112~~ \mu = (m_Z + m_t)/2} $ &  \\
\hline
\end{tabular}
\caption{Total rate comparison for $gu \rightarrow tZ$ at the LHC 14 TeV: aNLO vs NLO from Ref.~\cite{Li:2011ek}. The cross section
$\sigma_0$ is the default central value obtained using the central scale choice $\mu=m_Z + m_t$, i.e.~$\sigma_0 = \sigma(\mu = m_Z + m_t)$.
The scale dependence is obtained by varying $\mu$ up and down by a factor of 2, i.e. $(m_Z + m_t)/2 \leq \mu \leq 2 (m_Z + m_t)$. CTEQ6M NLO PDFs~\cite{Pumplin:2002vw} are used. }
\label{Comparison-Table}
\end{center}
\end{table}
These numbers are in very good agreement (within 2 per mille) with Ref.~\cite{Li:2011ek} and can be checked
in Table I and Fig. 6 respectively in that paper. They show that the soft-gluon approximation is excellent
for these processes. As also noted in Ref.~\cite{Kidonakis:2017mfy}, the agreement between aNLO and NLO is also very good for the $gc \rightarrow tZ$ channel.

\begin{figure}[htb!]
\begin{center}
\includegraphics[width=10.0cm]{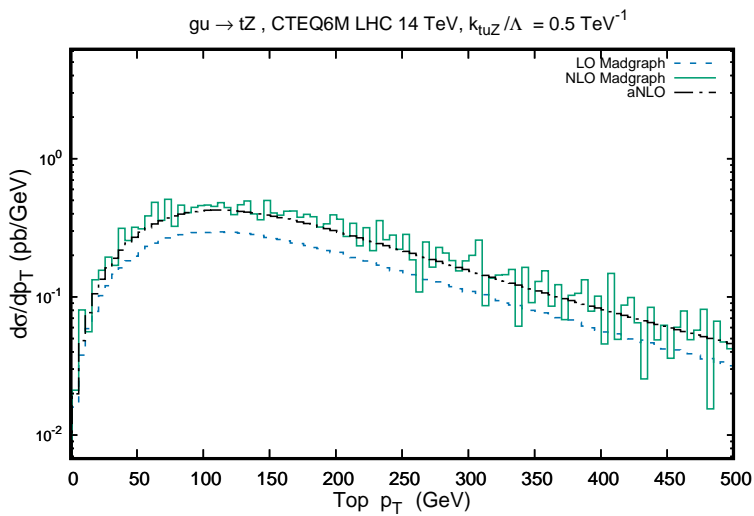}
\caption{aNLO vs NLO Madgraph top-quark $p_T$ distribution at the 14 TeV LHC. }
\label{aNLOvsNLO-top-pT}
\end{center}
\end{figure}

A second independent cross check for the $gu \rightarrow tZ$ channel was made by using
\textsc{MadGraph5\_aMC@NLO}~\cite{Alwall:2014hca} which provides both the total rate and the top-quark $p_T$ distribution.
We have used the FCNC Madgraph module described in Refs.\cite{Degrande:2014tta,Durieux:2014xla} which employs a general approach to top-quark
FCNC based on effective field theory. We fixed the parameters such that we could compare the
cross section relative to the tensor interaction term only in the Lagrangian.
We obtained the results illustrated in Fig.~\ref{aNLOvsNLO-top-pT} where the aNLO prediction is in very good agreement with the NLO calculation.

\begin{figure}[htb!]
\begin{center}
\includegraphics[width=10.0cm]{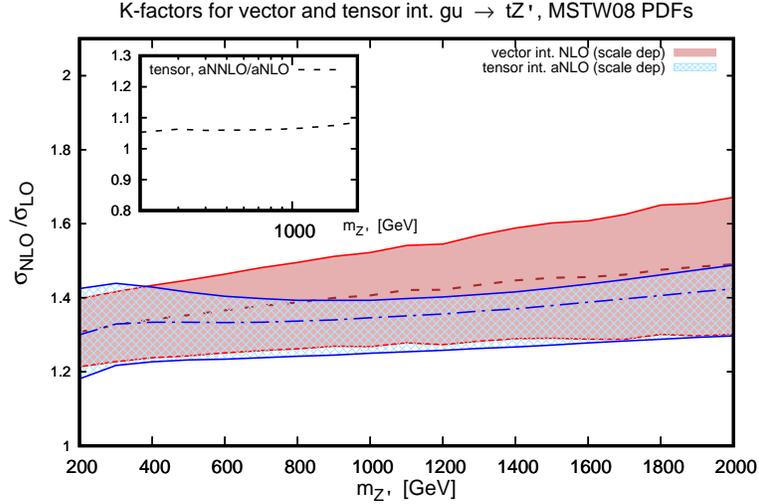}
\caption{NLO $K$-factors for top FCNC with vector interactions and the
aNLO $K$-factors for top FCNC with tensor interactions for the $gu\rightarrow tZ'$
channel at the LHC at 7 TeV. Scale variation refers
to $(m_{Z'}+m_t)/4\leq \mu \leq m_{Z'}+m_t$. The inset plot also shows the aNNLO $K$-factors.}
\label{K-fact-7tev}
\end{center}
\end{figure}

In the case of $tZ'$ production, our aNLO results have also been compared to the full NLO calculation at 7 TeV
LHC energy provided in Ref.~\cite{Adelman:2012py}. In particular, we compared $K$-factors. It is important to notice
that the Lagrangian used to obtain the results in Ref.~\cite{Adelman:2012py} only includes
vector interaction contributions, e.g.,
${\cal L}_{Z'}=(Q_{tU}/\sqrt{2}) \bar{U} \gamma^{\mu} g_R P_R t Z_{\mu}^\prime + ~h.c.$, where $Q_{tU}$
is a coupling factor, $g_R$ is a coupling constant, $P_R$ is the right-handed chiral projector, and $U$ denote the generic up-type quark.
In this study, we consider only tensor interactions (cf. Eq.~(\ref{Langrangian})). Moreover, the authors of Ref.~\cite{Adelman:2012py} have used different PDFs,
MSTW2008 \cite{Martin:2009iq}, and a different choice of central scale, $(m_{Z'}+m_t)/2$, than our choice of central scale, $m_{Z'}$.
Therefore, to make a valid comparison between $K$-factors from vector and tensor interactions, we have adopted their PDFs and scale choices
to make a comparison at 7 TeV. Because they use Run 1 LHC energies, the authors of Ref.~\cite{Adelman:2012py} only show results up to $m_{Z'}$ masses of 2000 GeV.
In Fig.~\ref{K-fact-7tev} we display NLO/LO $K$-factors relative to vector interactions and aNLO/LO $K$-factors relative to tensor interactions
for $\mu=(m_{Z'}+m_t)/2$, and a variation of that scale by a factor of two up and down for the NLO
and aNLO corrections at these scales relative to the central LO result at $\mu=(m_{Z'}+m_t)/2$.
We find that tensor interactions give $K$-factors at aNLO which are very similar in magnitude to those obtained by using
vector interactions, but the scale dependence for the tensorial case is found to be somewhat smaller than (but consistent with) the vector case.
We stress, however, that we do not expect exact agreement between the two cases due to the different Lagrangians involved.
In the inset plot of Fig.~\ref{K-fact-7tev} we also display the additional enhancements from the aNNLO corrections,
where in the numerator of aNNLO/aNLO we use NNLO PDFs and in the denominator we use NLO PDFs.

In conclusion, we have shown that for $tZ$ production the soft-gluon corrections account for the overwhelming majority of the complete corrections
and that the aNLO calculation is very trustworthy. This was already demonstrated for $tZ$ production in Ref.~\cite{Kidonakis:2017mfy} and it is
also consistent with the fact that the NLO soft-gluon corrections approximate very well the complete NLO corrections
for $t\gamma$~\cite{Forslund:2018qcp} production via anomalous couplings,
as well as for top-pair~\cite{NKttbar} and single-top~\cite{NKsingletop} production.

\subsection{Matching to the NLO theory at fixed order in QCD}

The formalism utilized in this study is expected to work equally well in the case of $tZ'$ production, because
it is essentially the same, the only difference being that the mass of the $Z'$ can have different values.
Indeed, after performing the aNLO and NLO calculations for $tZ'$ production for a variety of collider energies and $Z'$ masses, we observed that the aNLO and the exact NLO results differ by a few percent.
As expected, at large collider energies and large $m_{Z'}$ values, soft-gluon corrections account for the overwhelming majority of the QCD corrections, and the difference between the approximate and the exact NLO predictions is found to be very small.

To further improve our theoretical predictions, we match our aNNLO prediction to the exact NLO theory at fixed order
in QCD, and in the rest of this paper we show phenomenological results at NLO and aNNLO.
The NLO fixed order theory prediction for both the FCNC and the stringy inspired $tZ'$ production is obtained with
\textsc{MadGraph5\_aMC@NLO-v2.7.2}, which we have used to calculate both the total rate and the top-quark $p_T$ distributions.
The approximate aNNLO theory prediction is obtained by matching to the NLO as follows:
\begin{equation}
\sigma_{aNNLO} = \left[\hat{\sigma}_{LO} +  \hat{\sigma}_{NLO} + \hat{\sigma}_{aNNLO} \right]_{ij} \otimes f_i^{NNLO} \otimes f_j^{NNLO}\,,
\end{equation}
where the soft-gluon contributions from the aNNLO hard scattering are added on top of the fixed-order NLO.
The matching procedure ensures a better control of kinematic regions of the phase space where soft-gluons are less dominant.

\subsection{FCNC $Z'$s: $gu \rightarrow tZ'$ and $gc \rightarrow tZ'$}

We first study $tZ'$ production via FCNC interactions with anomalous couplings.
The partonic processes involved are $gu \rightarrow tZ'$ and $gc \rightarrow tZ'$,
where the $Z'$ anomalously couples to the top quark and the $u$ and $c$ quarks through the
flavor-changing coefficients $k_{tuZ'}/\Lambda$ and $k_{tcZ'}/\Lambda$, respectively.
The scale $\Lambda$ is set equal to ten times the top quark mass $m_t$ and the couplings
$k_{tuZ'}$ and $k_{tcZ'}$ are considered as parameters of the theory. As a case study we
select $k_{tuZ'}=k_{tcZ'}=0.1$. Thus, in our results below we
set $k_{tuZ'}/\Lambda=k_{tcZ'}/\Lambda=0.01/m_t$. We also set $m_t=172.5$ GeV.
Recent experimental searches for and phenomenological studies of FCNC interactions
between the top quark and a $Z$ boson can be found in
Refs.~\cite{CMS:2016bss,Aad:2015uza,Khachatryan:2015att,Khanpour:2014xla,Hou:2017ozb}.

\begin{figure}[th!]
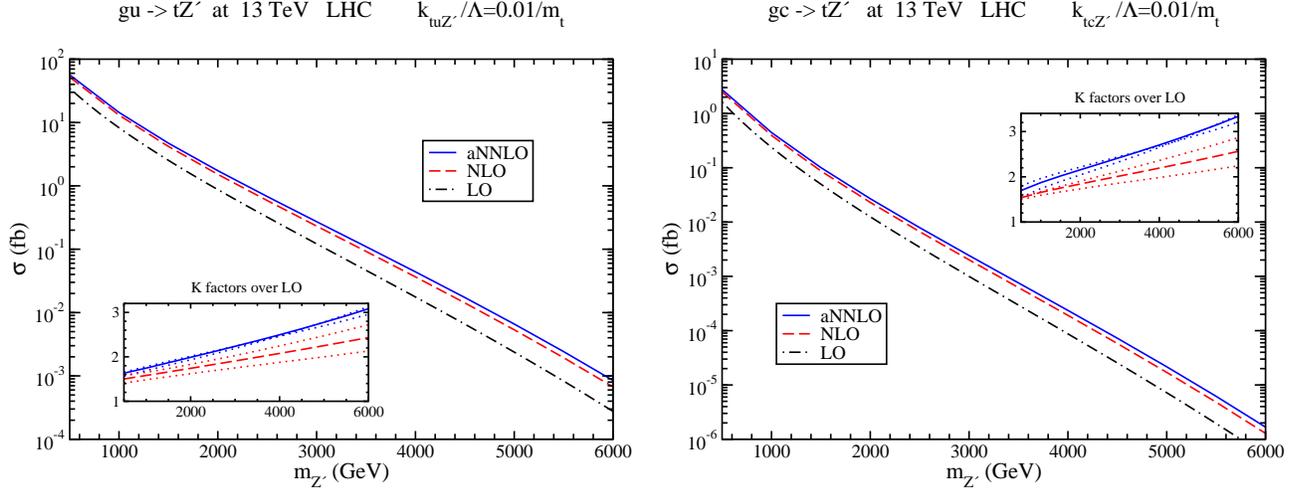

\begin{center}
\includegraphics[width=8.3cm]{gutZp13lhcnewplot.eps}
\hspace{1mm}
\includegraphics[width=8.3cm]{gctZp13lhcnewplot.eps}
\caption{Total cross sections at 13 TeV LHC energy for (left) $gu\rightarrow tZ'$
with anomalous $t$-$u$-$Z'$ coupling and (right) $gc\rightarrow tZ'$ with anomalous $t$-$c$-$Z'$ coupling.
The inset plots display $K$-factors.
Here CT14NNLO PDFs are used for the LO, NLO, and aNNLO
calculations to show the enhancement due to hard-scattering contributions.}
\label{gutZp13}
\end{center}
\end{figure}

\begin{figure}[th!]
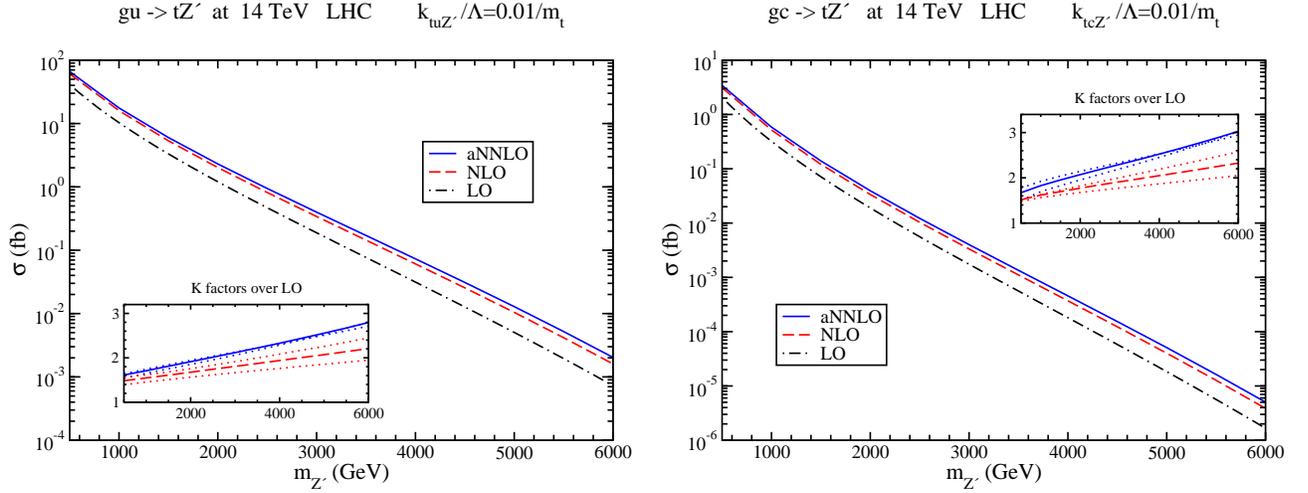

\begin{center}
\includegraphics[width=8.3cm]{gutZp14lhcnewplot.eps}
\hspace{1mm}
\includegraphics[width=8.3cm]{gctZp14lhcnewplot.eps}
\caption{Total cross sections at 14 TeV LHC energy for the (left) $gu\rightarrow tZ'$ and
(right) $gc\rightarrow tZ'$ processes with anomalous couplings. The inset plots display $K$-factors.
Here CT14NNLO PDFs are used for the LO, NLO, and aNNLO calculations to show the enhancement due to hard-scattering contributions.}
\label{gutZp14}
\end{center}
\end{figure}

\begin{figure}
\begin{center}
\includegraphics[width=8.3cm]{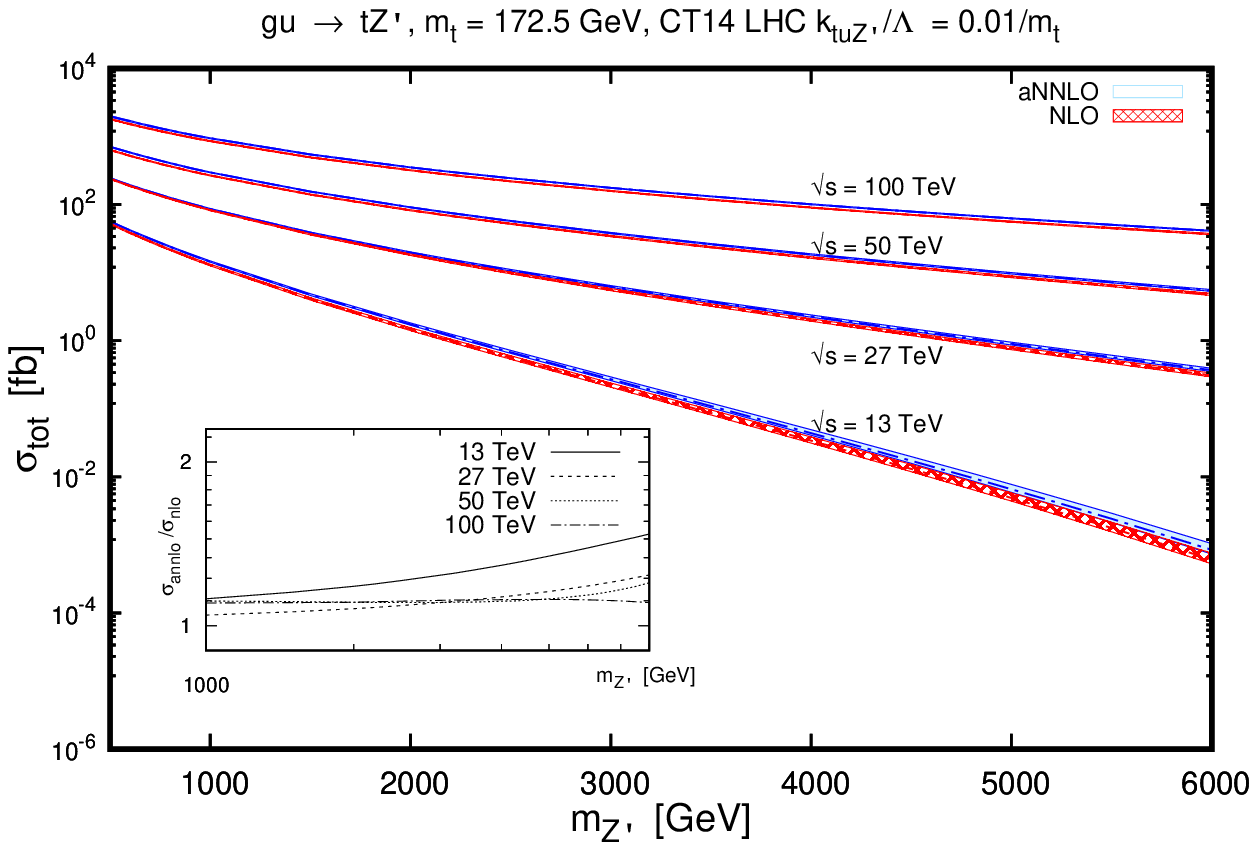}
\hspace{1mm}
\includegraphics[width=8.3cm]{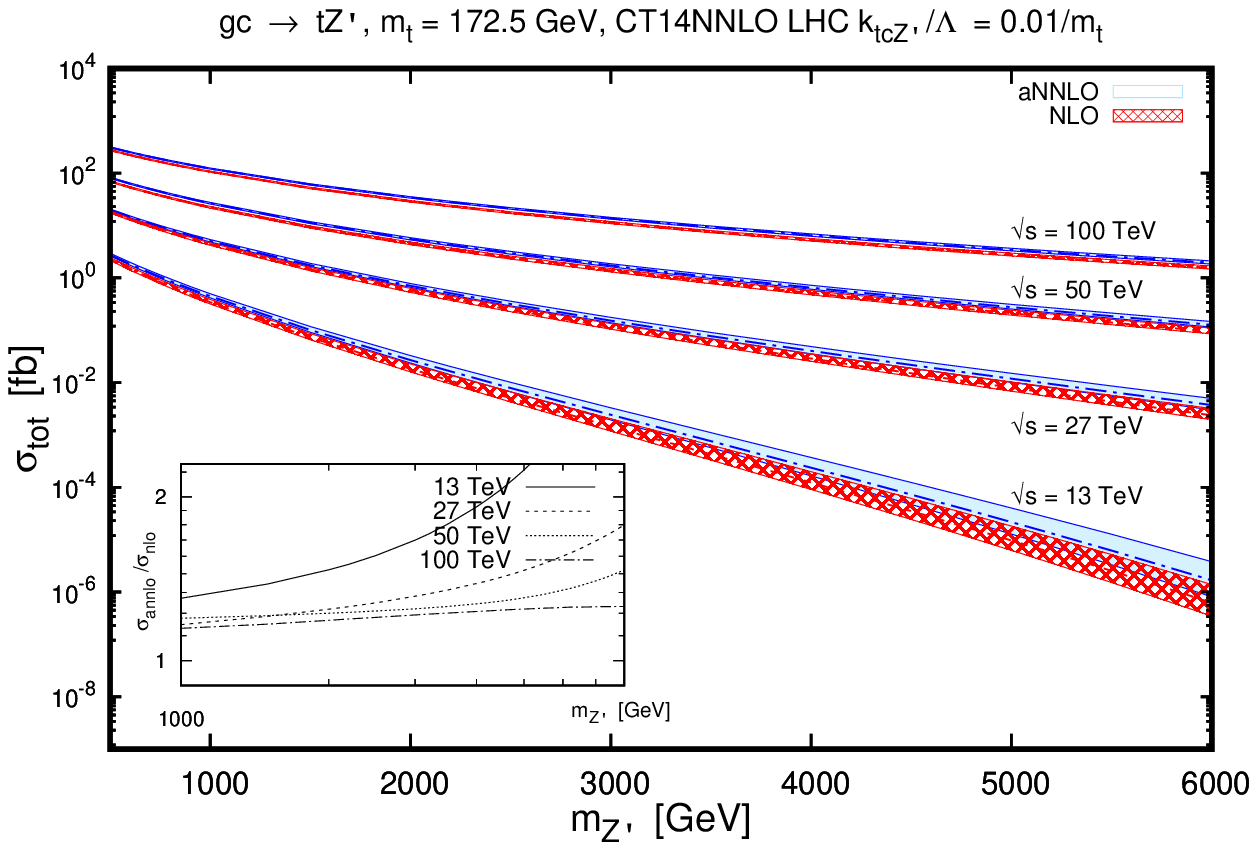}
\caption{Total cross sections for the (left) $gu \rightarrow tZ'$ and (right) $gc\rightarrow tZ'$ processes
with anomalous couplings. The plots show results including CT14 PDF uncertainties for several center-of-mass energies
of the $pp$ collision as a function of $Z'$ mass. The aNNLO cross section is obtained with CT14NNLO PDFs
while the NLO with CT14NLO. The CT14 PDF uncertainties are at the 68\% C.L.. The inset plots show the $\sigma_{aNNLO}/\sigma_{NLO}$ $K$-factors.}
\label{gutZpscan}
\end{center}
\end{figure}

\begin{figure}[th!]
\begin{center}
\includegraphics[width=8.3cm]{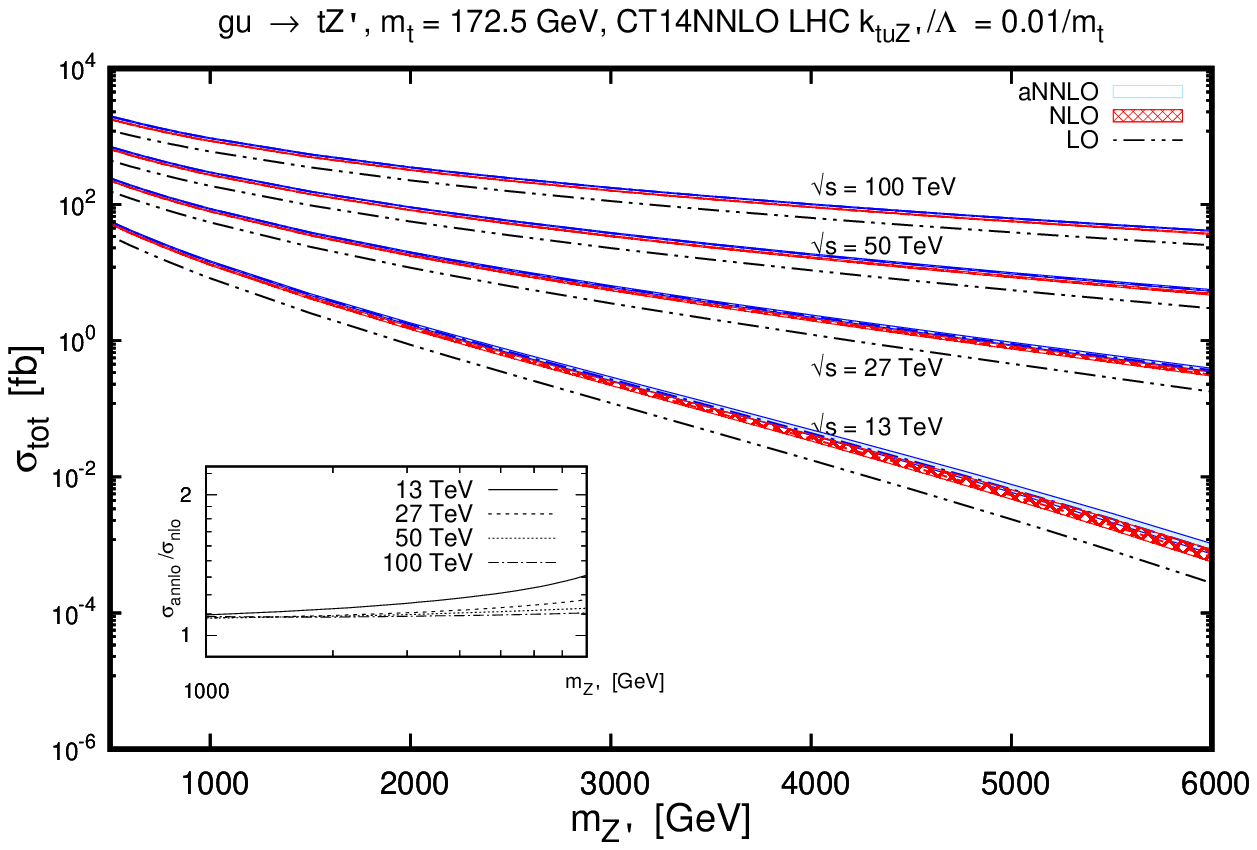}
\hspace{1mm}
\includegraphics[width=8.3cm]{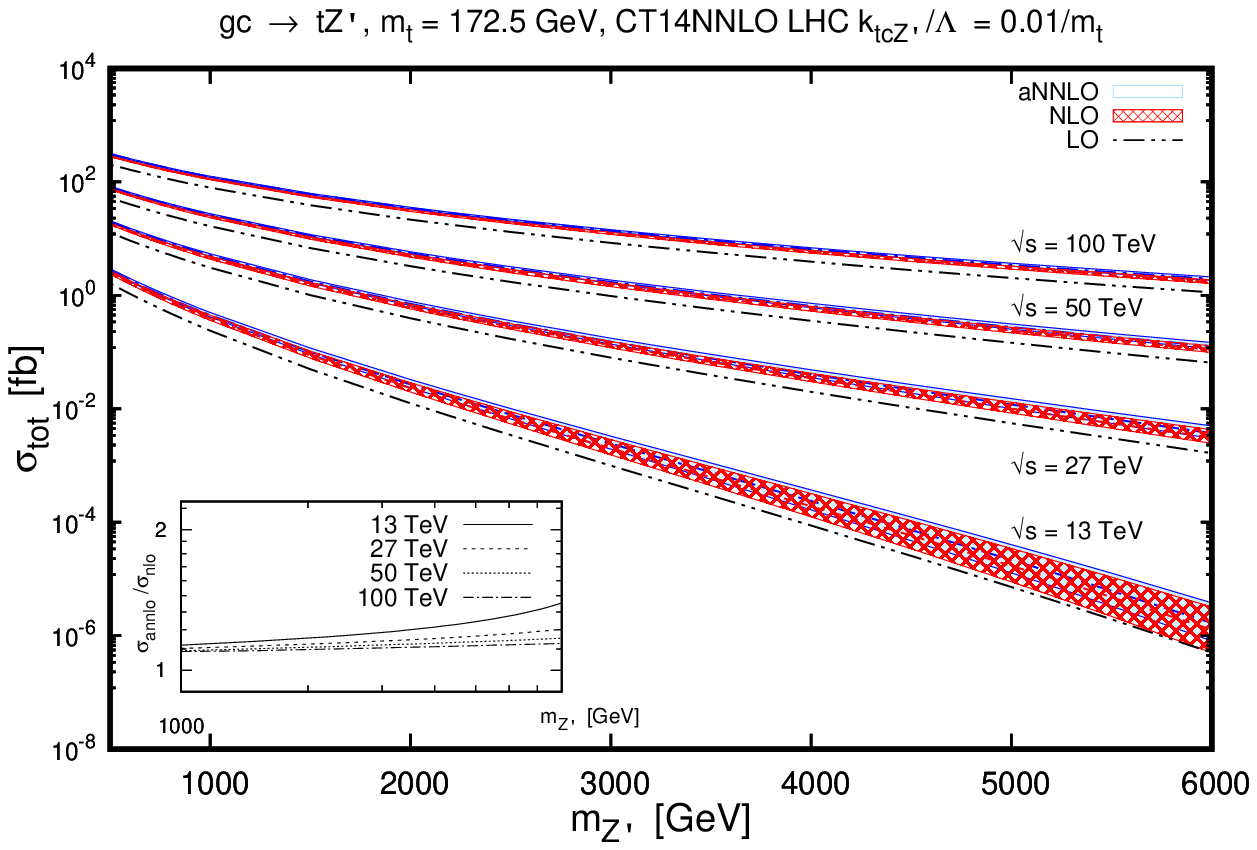}
\caption{Total cross sections for the $gu \rightarrow tZ'$ (left) and the $gc \rightarrow tZ'$ (right) processes with anomalous couplings.
The plots show results with CT14 NNLO PDF uncertainties at 68\% C.L. for several center-of-mass energies of the $pp$ collision as a function of
$m_{Z'}$.}
\label{MZpscan-ct14nnlo}
\end{center}
\end{figure}

\begin{figure}
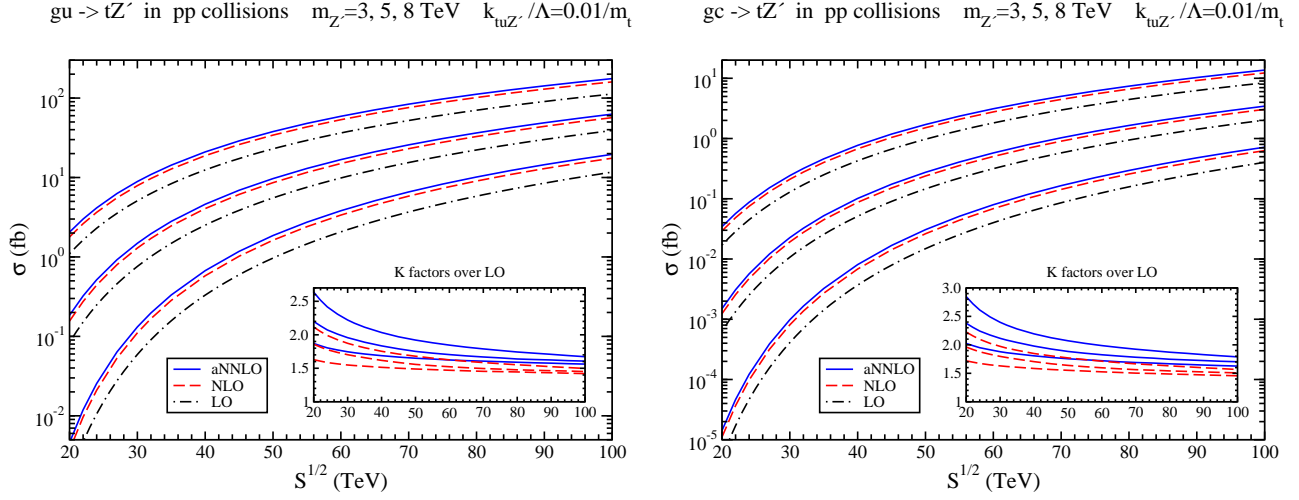

\begin{center}
\includegraphics[width=8.3cm]{gutZprootSnewplot.eps}
\hspace{1mm}
\includegraphics[width=8.3cm]{gctZprootSnewplot.eps}
\caption{Total cross sections for the (left) $gu\rightarrow tZ'$ and (right)
$gc\rightarrow tZ'$ processes with anomalous couplings. The plots show results
as a function of collider energy for three choices of $Z'$ mass, 3, 5, and 8 TeV.
The inset plots display $K$-factors. CT14NNLO PDFs are used.}
\label{gutZprootS}
\end{center}
\end{figure}

\begin{figure}
\begin{center}
\includegraphics[width=8.4cm]{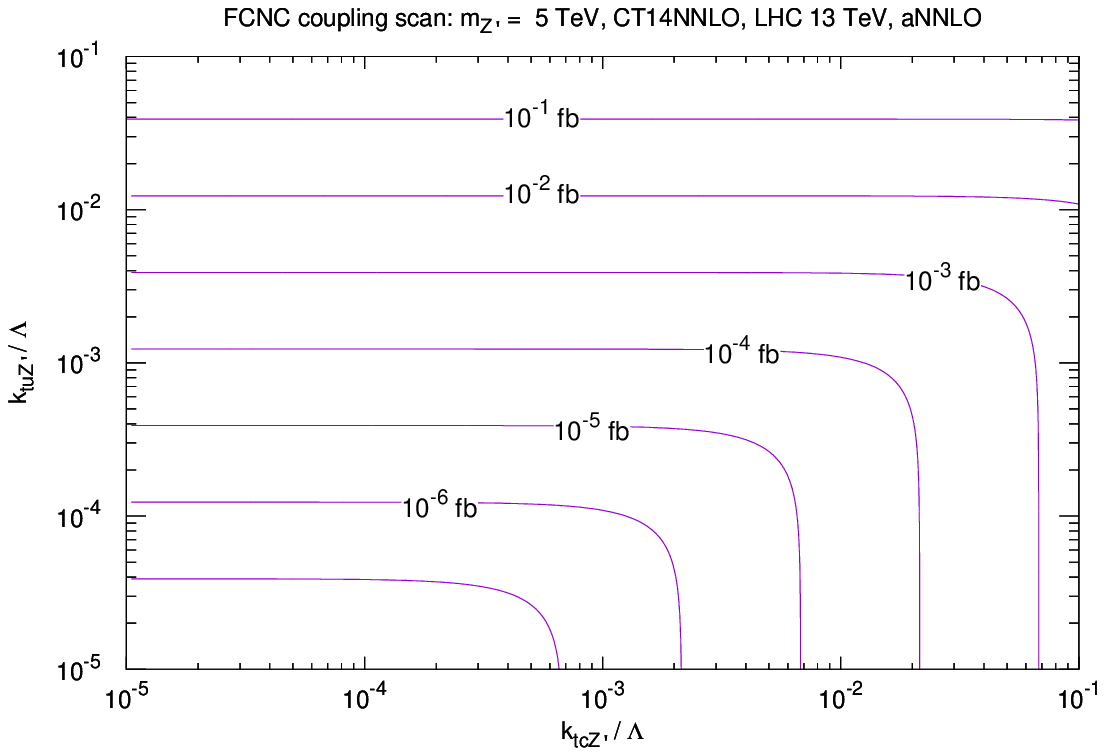}
\includegraphics[width=8.4cm]{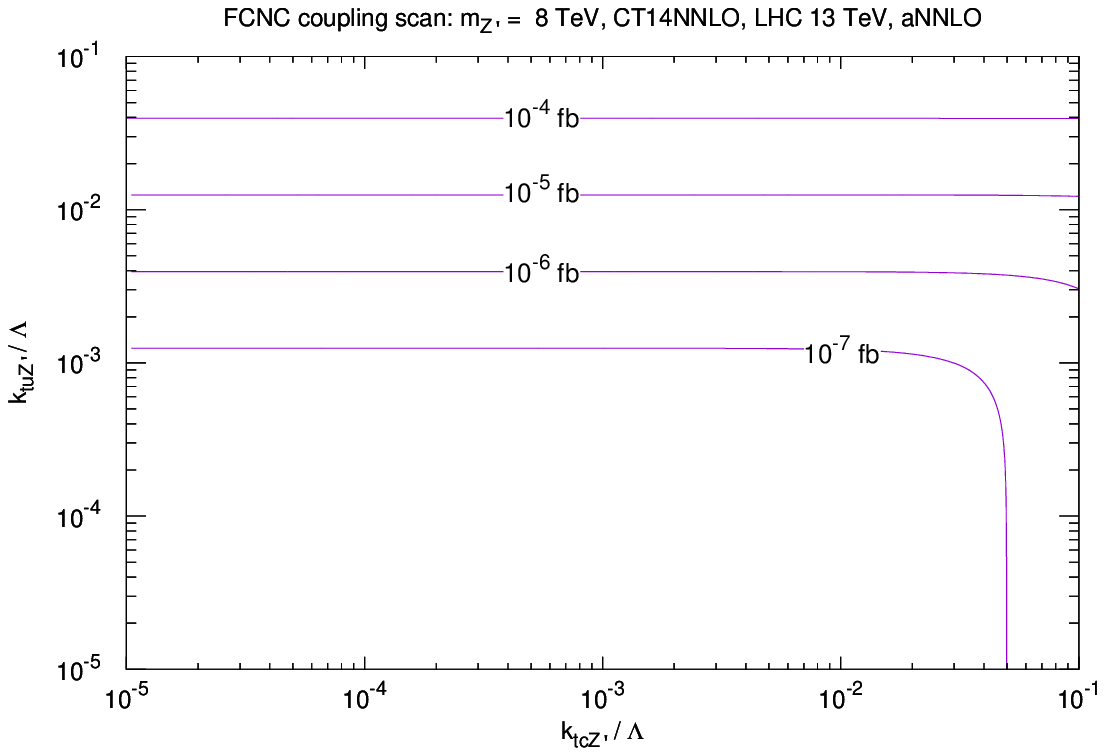}
\\
\includegraphics[width=8.4cm]{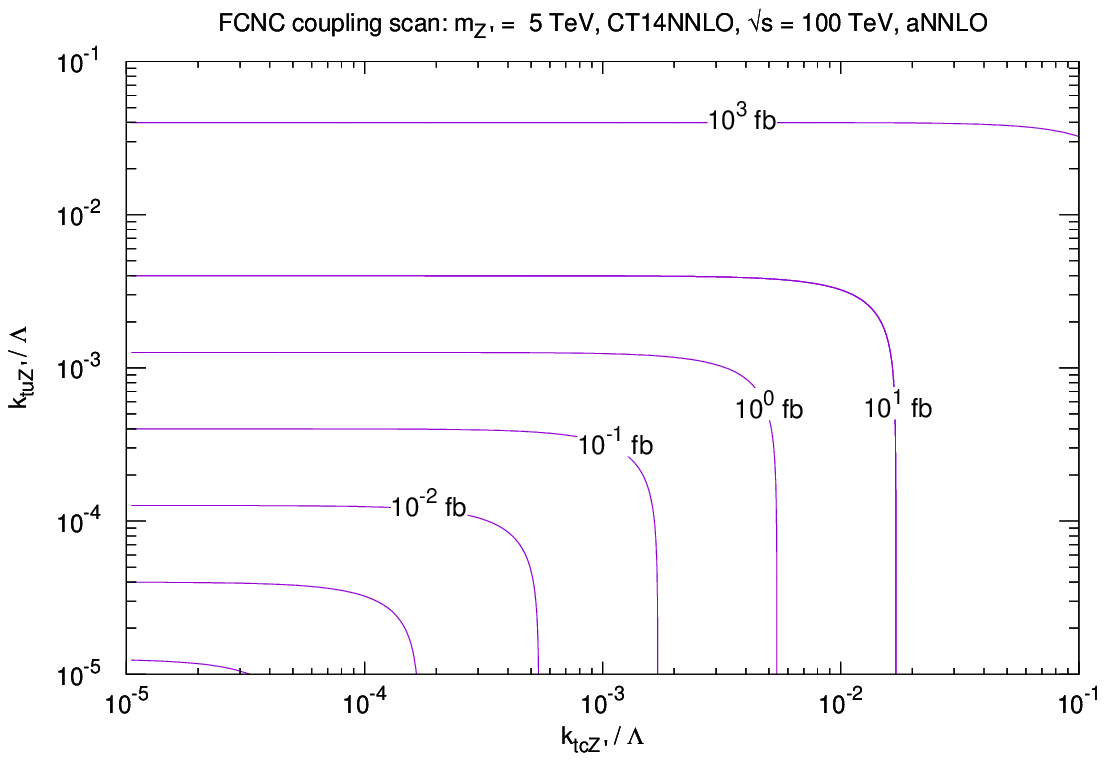}
\includegraphics[width=8.4cm]{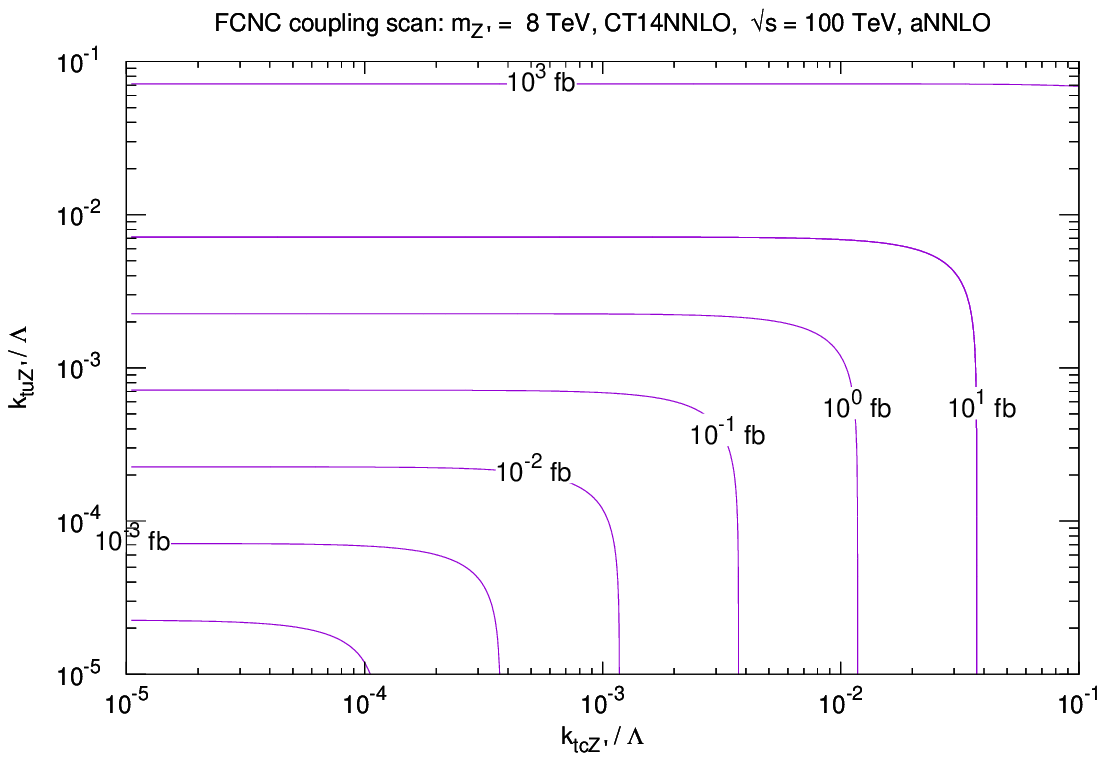}
\caption{Total cross sections for the $(gc+gu)\rightarrow tZ'$ process in a 2D contour plot.
The insets show aNNLO results as a function of the anomalous couplings $k_{tuZ'}/\Lambda$ and
$k_{tcZ'}/\Lambda$ (given in units of inverse $m_t$), in $pp$ collisions at $\sqrt{S}$=13 and 100 TeV. CT14NNLO PDFs are used.}
\label{coupling-scan}
\end{center}
\end{figure}

We explore cross sections at 13 and 14 TeV LHC energies for a large
range of $Z'$ masses, and also explore the cross sections as functions of
$pp$ collider energy for future colliders. The theory predictions in this
case are obtained by using the CT14 PDFs~\cite{Dulat:2015mca} which lead
to the numerical results illustrated in Figs.~\ref{gutZp13}-\ref{K-fact-scaleunc}.
In this case, PDF induced uncertainties are calculated at the 68\% confidence level (C.L.)
(see Appendix \ref{AppA} for a discussion on PDF uncertainties).

The initial-state parton combinations $g(x_1)u(x_2) + g(x_2)u(x_1)$ and
$g(x_1)c(x_2) + g(x_2)c(x_1)$  are probed in various kinematic regions
depending on the collider center-of-mass energy and on the mass of the $Z'$.
At $\sqrt{S}= 13$ TeV and $1\lesssim M_{Z'}\lesssim 8$ TeV, one probes large
$x$ values $x\geq 0.1$ where the current PDFs are not well constrained and
their uncertainties are large. At higher collider energies $\sqrt{S}= 100$ TeV,
one probes $10^{-4} \lesssim x \lesssim 0.1$ for $ M_{Z'}\approx 1$ TeV,
and $0.01 \lesssim x \lesssim 0.1$ for $M_{Z'}\approx 8$ TeV.

The total cross sections at collider energies of 13 TeV are illustrated
in Fig.~\ref{gutZp13} where we show the theory predictions
at LO, NLO, and aNNLO for the process $gu\rightarrow tZ'$ with anomalous $k_{tuZ'}$ coupling,
and the process $gc\rightarrow tZ'$ with anomalous $k_{tcZ'}$ coupling, as functions of $Z'$ mass.
Here CT14NNLO PDFs are used for the LO, NLO, and aNNLO calculations to show soft-gluon enhancements
in the hard-scattering contributions with respect to the Born cross section.
The factorization and renormalization scales are equal and set to $\mu=m_{Z'}$.
We observe a very strong dependence of the cross section on the $Z'$ mass. The cross section drops
over many orders of magnitude as the $Z'$ mass varies from 1 TeV to 6 TeV.
The cross section for $gc \rightarrow tZ'$ is significantly smaller than for $gu\rightarrow tZ'$.
The inset plots show the NLO/LO and aNNLO/LO $K$-factors with scale uncertainty bands which are obtained
by varying $\mu$ in the interval $[1/2\mu , 2\mu]$ in the numerator. The $K$-factors are large and increase with larger $Z'$
masses, as expected. The NLO corrections are large and furthermore the additional aNNLO corrections are very significant.
We also provide numerical values for the $gu\rightarrow tZ'$ cross section and $K$-factors in Table \ref{tZ'table-gu}
of Appendix \ref{AppC}.

The corresponding results at 14 TeV energy are shown in Fig.~\ref{gutZp14}. The cross sections are of course larger than at 13 TeV,
but the dependence on the $Z'$ mass and the size of the corrections are very similar.

In Fig.~\ref{gutZpscan} we show the total cross sections at NLO and aNNLO for
the processes $gu\rightarrow tZ'$ and $gc\rightarrow tZ'$ at 13, 27, 50, and 100 TeV collider
energies together with CT14 PDF uncertainties evaluated at the 68\% confidence level (C.L.).
In this case, the aNNLO total cross sections are obtained with CT14NNLO PDFs,
while the NLO's are obtained with CT14NLO.
The inset plots show the $\sigma_{aNNLO}/\sigma_{NLO}$ $K$-factors.
We note that the $\sigma_{aNNLO}/\sigma_{LO}$ $K$-factors are not shown here because there are no CT14 PDFs at LO.
We observe that the $\sigma_{aNNLO}/\sigma_{NLO}$ $K$-factors provide large corrections
for large values of $m_{Z'}$, and the corrections decrease as the collider energy increases.
The induced PDF uncertainty of both $gu$ and $gc$ channels is larger at lower collider
energy and high $m_{Z'}$ where PDFs are weakly constrained.
Figure~\ref{MZpscan-ct14nnlo} shows total cross section predictions at 13, 27, 50, and 100 TeV collider energies using CT14NNLO PDFs at all orders for FCNC $tZ'$ production to show the enhancement due to soft gluons in the perturbative series.

The behavior of the cross section with collider energy is illustrated in Fig.~\ref{gutZprootS},
where we show results at LO, NLO, and aNNLO for the $gu$ and $gc$ channels as functions of
the collider energy up to 100 TeV for three choices of $Z'$ mass, $m_{Z'}=3$, 5, and 8 TeV.
Here, the LO, NLO, and aNNLO cross sections are obtained with CT14NNLO PDFs to show enhancement in the hard scattering
due to soft gluon corrections.
The cross sections are smaller for larger $Z'$ masses due to phase-space suppression.
The inset plots show the NLO/LO and aNNLO/LO $K$-factors. As expected, the $K$-factors are larger at smaller energies
and also for higher $Z'$ masses, since we are then closer to threshold.

In the case of $tZ'$ production with FCNC couplings, the anomalous
couplings entering both channels of the cross section are considered as
free parameters. We have therefore performed a two dimensional scan to assess the sensitivity of the cross section.
In Fig.~\ref{coupling-scan} we show a case study in which we plot
aNNLO total cross sections as functions of the couplings $k_{tuZ'}/\Lambda$ and $k_{tcZ'}/\Lambda$,
at a collider energies of 13 and 100 TeV, for different values of $m_{Z'}$.
We notice that if we let both couplings to vary in $10^{-5} \leq k/\Lambda\leq 0.1$ TeV$^{-1}$,
the cross section spans several orders of magnitude.
The cross section suppression is larger for larger values of $m_{Z'}$.

\subsubsection{Top-quark $p_T$ distributions for FCNC $Z'$s }

\begin{figure}
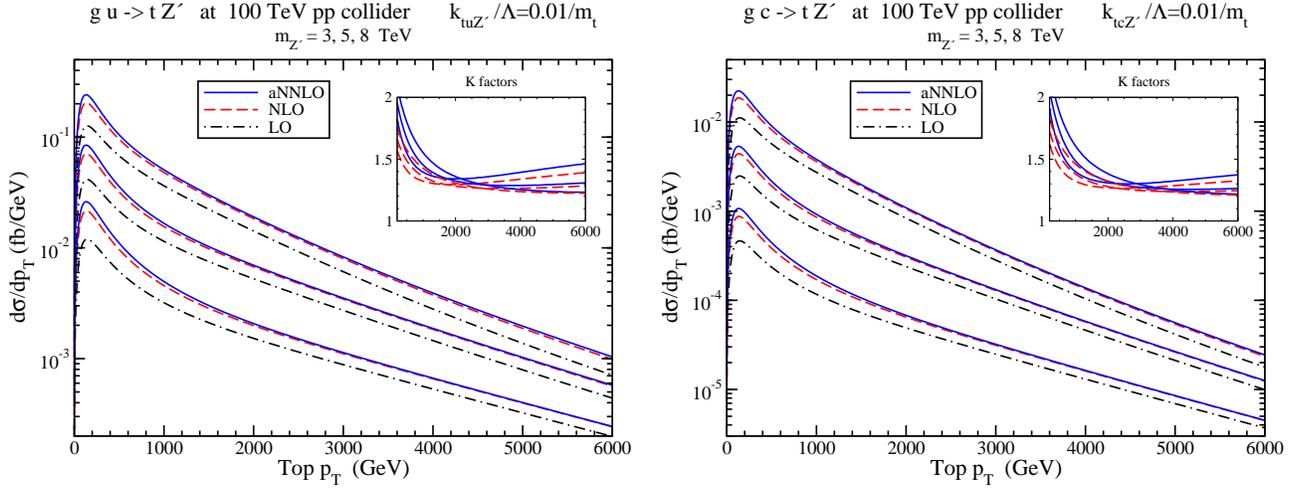

\begin{center}
\includegraphics[width=8.3cm]{pttopgutZp100tevnewplot.eps}
\hspace{1mm}
\includegraphics[width=8.3cm]{pttopgctZp100tevnewplot.eps}
\caption{Top-quark $p_T$ distributions for the (left) $gu \rightarrow tZ'$ and (right)
$gc\rightarrow tZ'$ processes with anomalous couplings for $m_{Z'}=3$, 5, and 8 TeV at 100 TeV $pp$ collider energy.
Inset plots: NLO/LO and aNNLO/LO K-factors. CT14NNLO PDFs are used.}
\label{ptgutZp100}
\end{center}
\end{figure}

It is interesting to study kinematic distributions such as the top-quark $p_T$ differential distribution, $d\sigma/dp_T$,
and how $Z's$ of different masses affect the $p_T$ suppression in various kinematic ranges.
We illustrate the top-quark $p_T$ distributions, calculated by a numerical
integration of the double-differential distribution, in Fig.~\ref{ptgutZp100}. Results for
the $gu \rightarrow tZ'$ and $gc \rightarrow tZ'$ processes at a collider energy of 100 TeV
are shown at LO, NLO, and aNNLO, obtained with CT14NNLO PDFs,
for three choices of the $Z'$ mass of 3, 5, and 8 TeV.

The NLO corrections are large and furthermore the additional aNNLO corrections are important.
The $p_T$ distributions decrease quickly as $m_{Z'}$ is increased, but they are non-negligible even for large $Z'$ masses, indicating that the number of events predicted by these models can be validated at the high-luminosity FCC or SppC colliders. The $K$-factors, shown in the inset plots, are significant and their value depends on $m_{Z'}$ and on the phase-space supression.

\subsubsection{Cross section and PDF correlations}

Next, we explore the extent of correlation between the PDFs and the aNNLO cross
section for these processes in $pp$ collisions at $\sqrt{S}$=13 and 100 TeV.
PDF correlations are important because they give us information about the kinematic region in which PDFs are
probed and for example, they give us indication of the impact of the gluon at different values of the momentum fraction $x$.
In order to set tighter constraints on $Z'$s models it is important to understand how PDF uncertainties
come into play and how to improve their precision through dedicated QCD global analyses.

\begin{figure}[tbh!]
\begin{center}
\includegraphics[width=8.3cm]{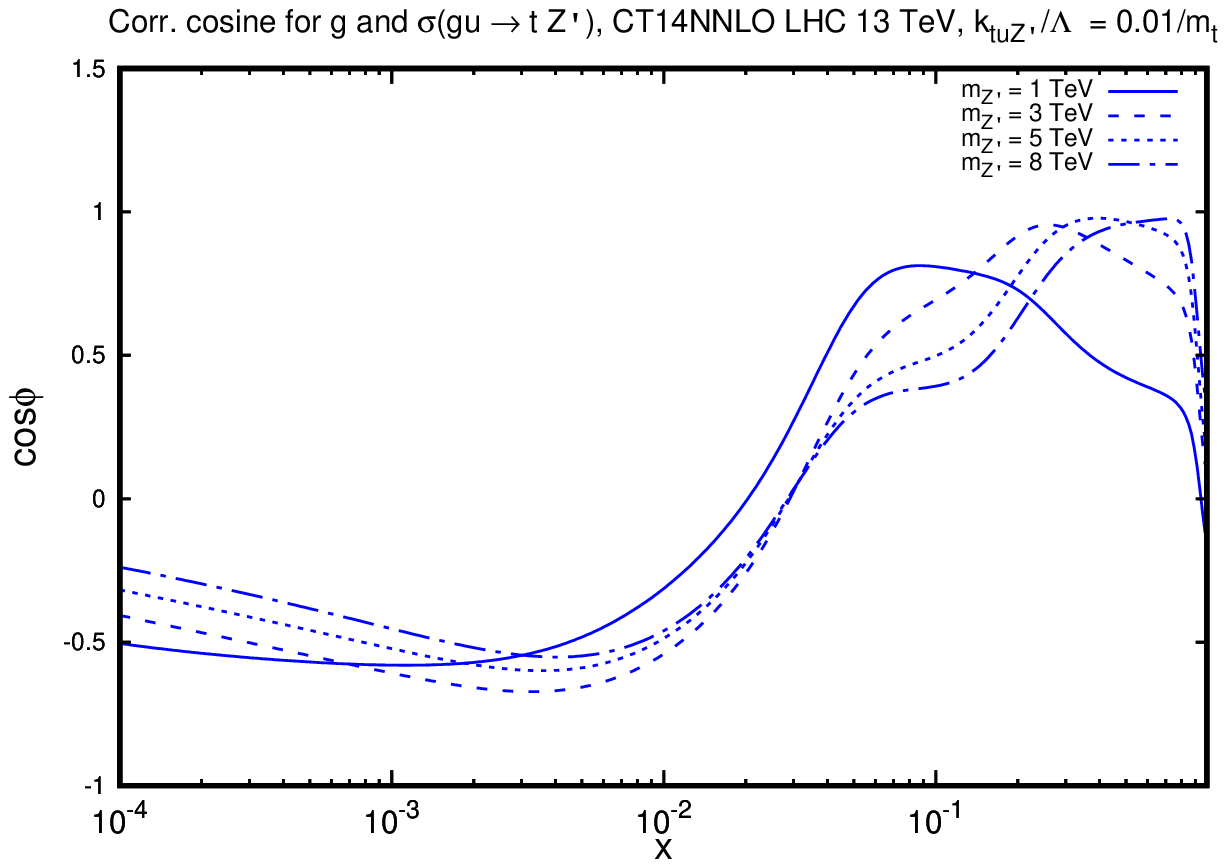}
\hspace{1mm}
\includegraphics[width=8.3cm]{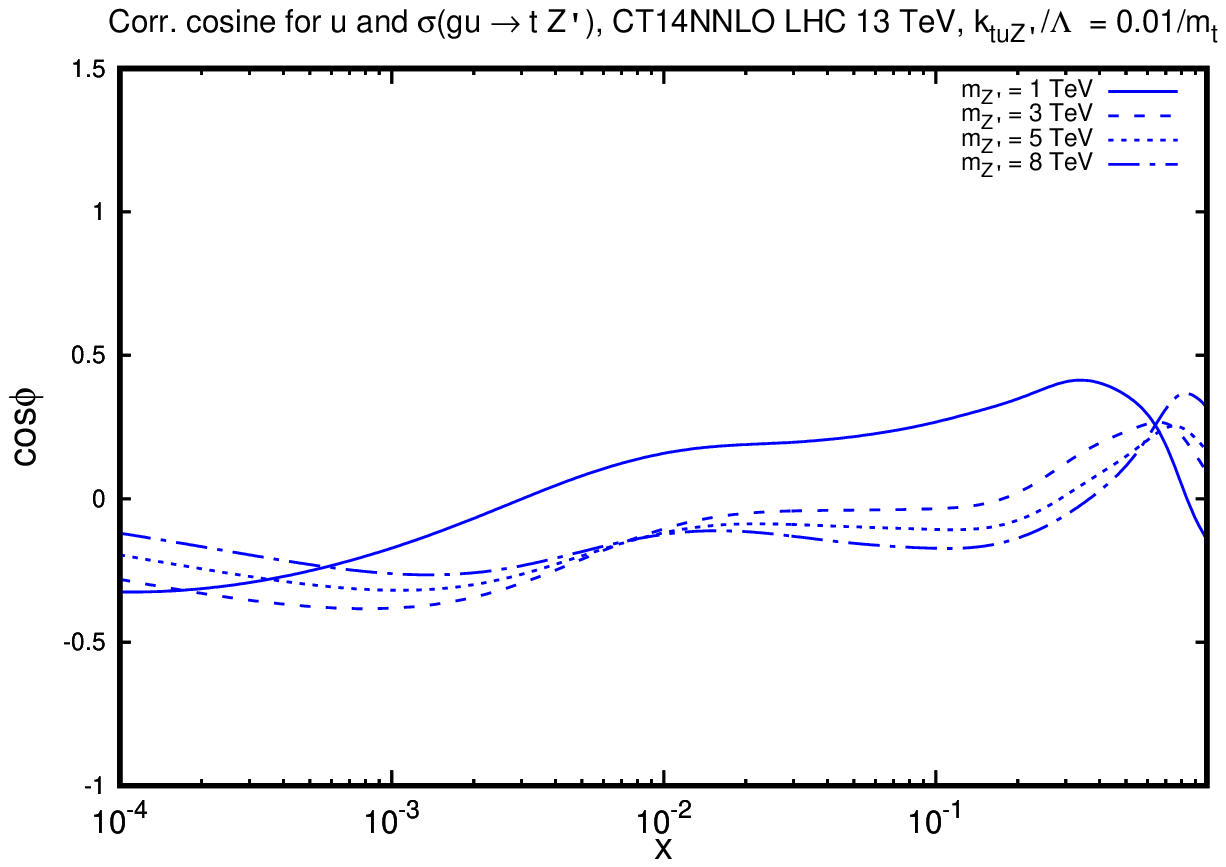}
\\
\includegraphics[width=8.3cm]{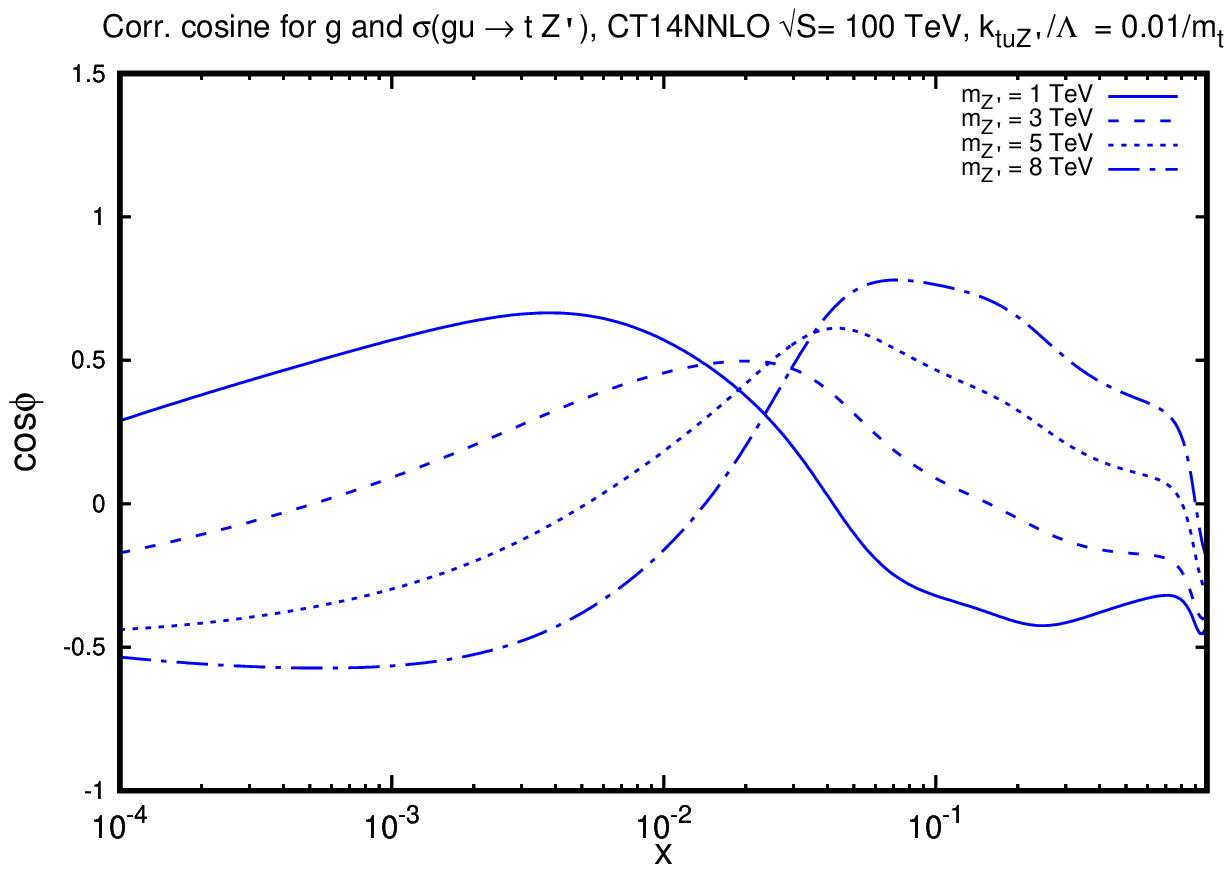}
\hspace{1mm}
\includegraphics[width=8.3cm]{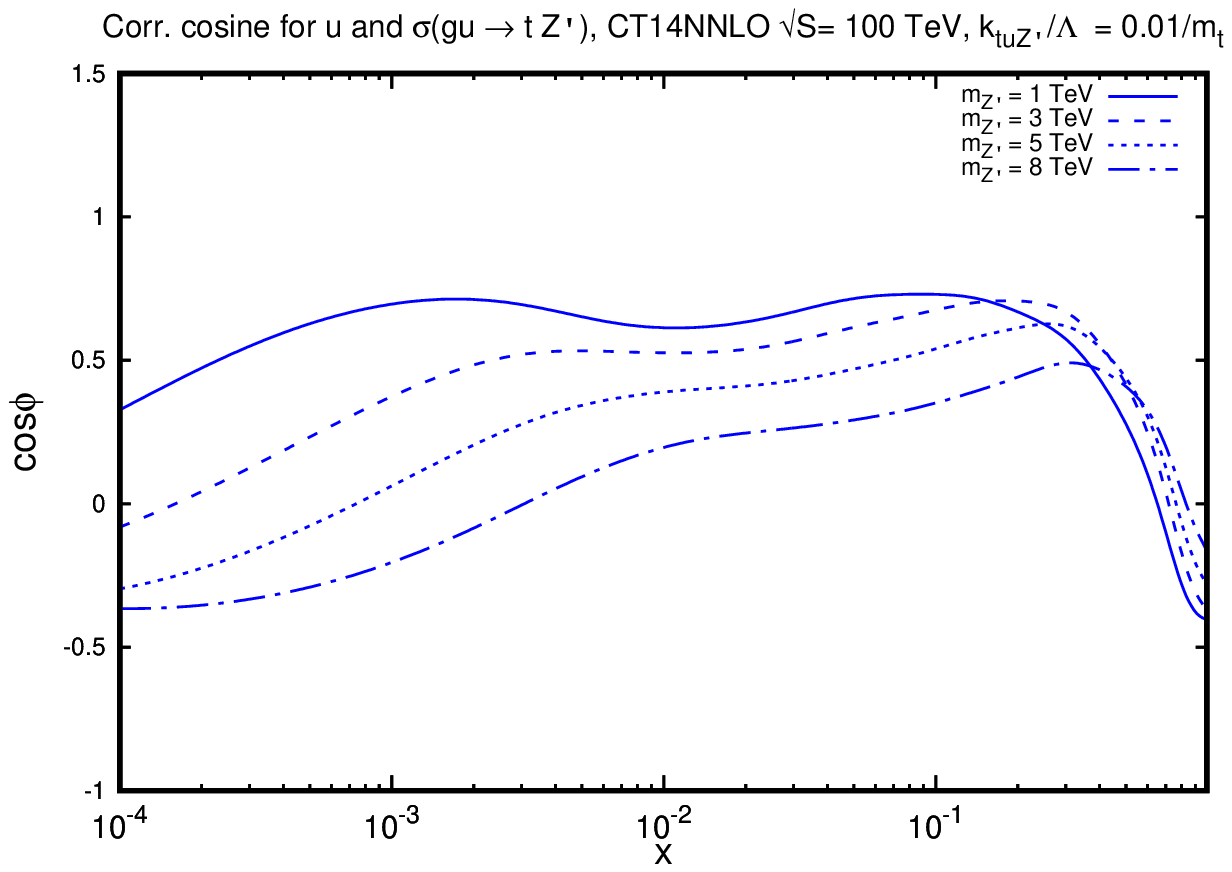}
\caption{Correlation cosine at $\sqrt{S}$=13 and 100 TeV between $\sigma_{gu\rightarrow tZ'}$ and
the gluon as a function of $x_{\textrm{gluon}}$ (left column) and $\sigma_{gu\rightarrow tZ'}$ and
the up-quark as a function of $x_{\textrm{up}}$ (right column). The four panels show aNNLO results rescaled at
the 68\% C.L. for different values of the $Z^\prime$ mass.}
\label{corr-cos-sgu-sgc}
\end{center}
\end{figure}

In particular, in Fig.~\ref{corr-cos-sgu-sgc} we show the correlation cosine
between the gluon (and the $u$ quark) and the total cross section for the
$gu \rightarrow t Z'$ process as a function of the momentum fraction $x$ at the
68\% CL at $\sqrt{S}=$ 13 and 100 TeV. We have chosen the $gu$ channel as it
provides the dominant contribution. The definition of the correlation cosine
between two quantities determined within the Hessian method is given in
Appendix~\ref{AppB}. At collider energies of 13 TeV, we observe a strong
correlation ($\cos{\phi}\geq 0.8$) between the gluon and the $gu \rightarrow t Z'$
cross section at large $x\geq 0.1$ as expected, and the correlation peak shifts
towards larger $x$ values for larger $m_{Z'}$. Anti-correlation of approximately
50\% in the $10^{-4} \leq x\leq 10^{-2}$ interval is also observed.
The correlation between the $u$ quark and the cross section is much milder and
less than 50\% at very large $x$. These patterns change as we move to higher
collider energies, where for the gluon the correlation peak for each value of $m_{Z'}$
is shifted to lower $x$-values, while for the $u$ quark correlations are slightly more pronounced.

\begin{figure}[tbh!]
\begin{center}
\includegraphics[width=8.3cm]{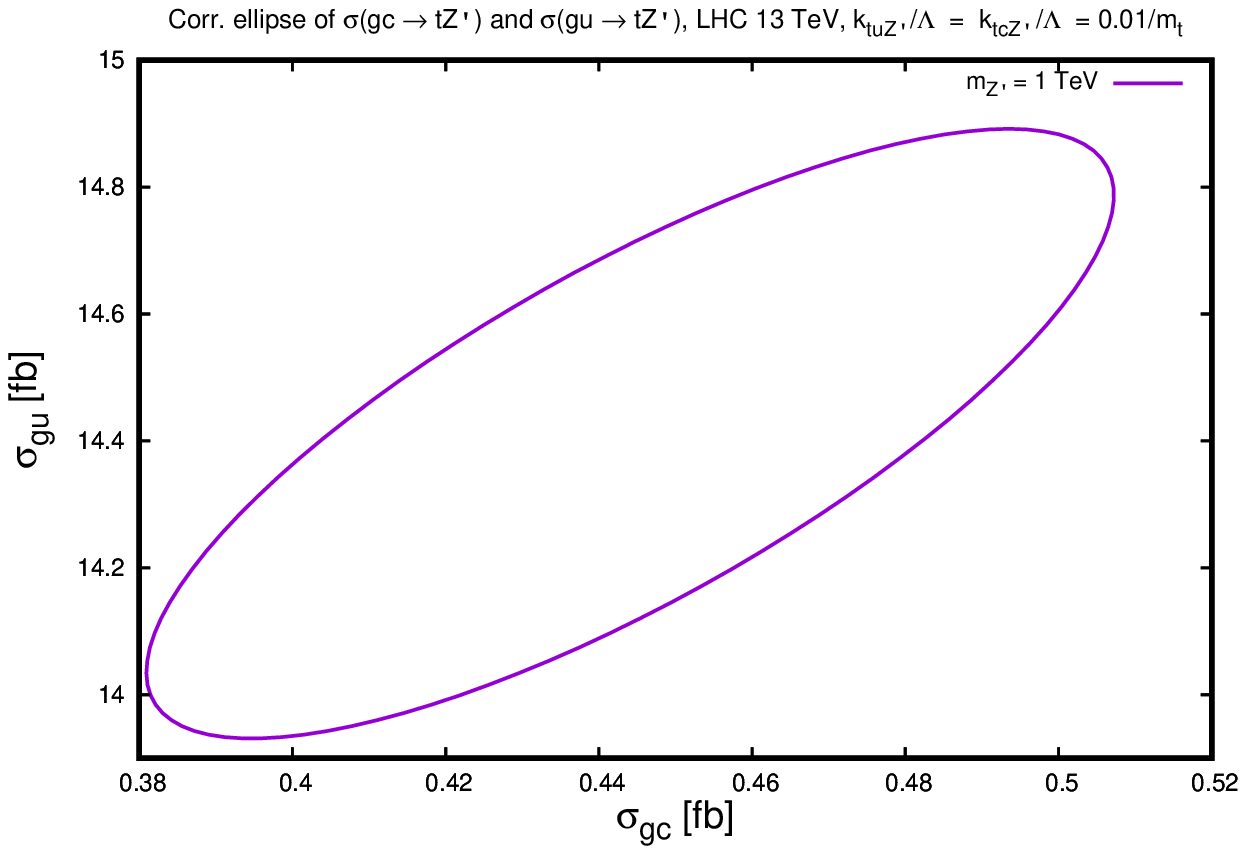}
\hspace{1mm}
\includegraphics[width=8.3cm]{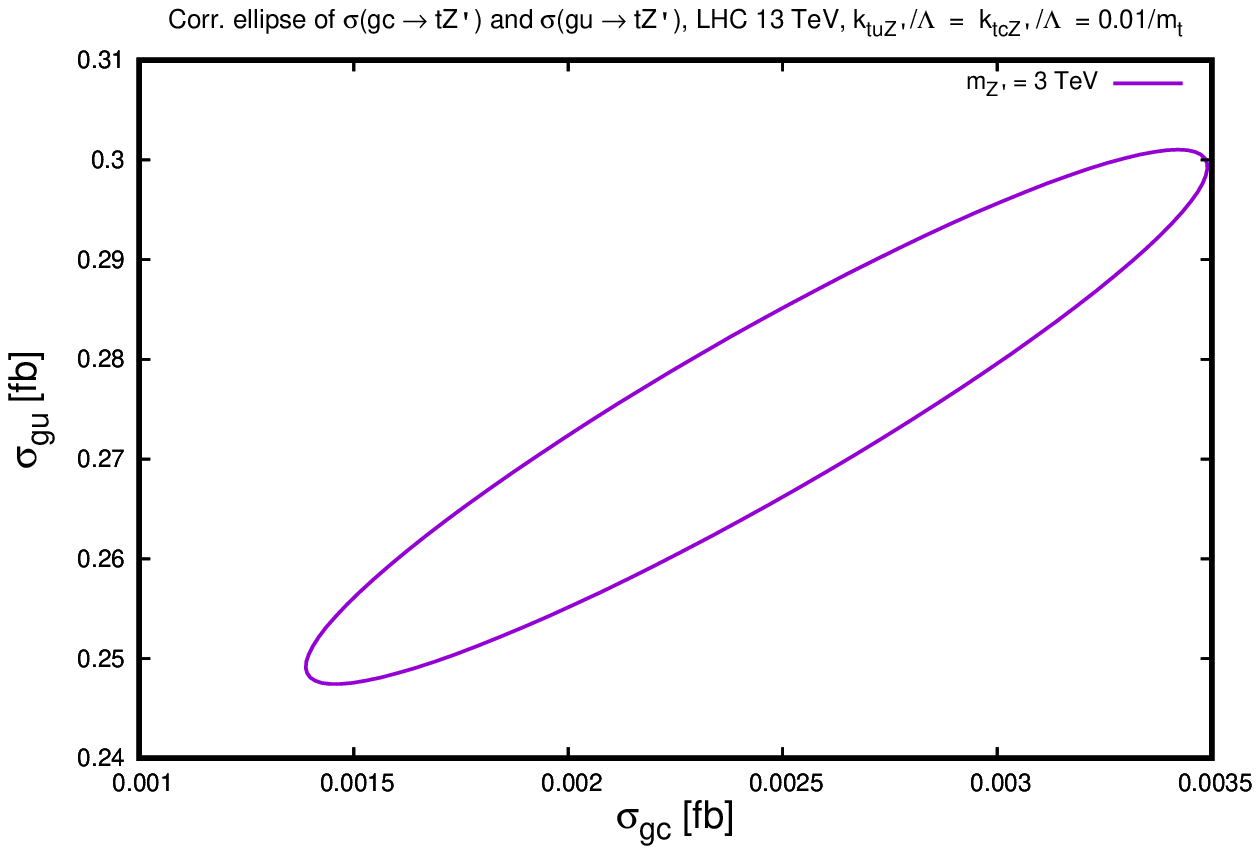}\\
\includegraphics[width=8.3cm]{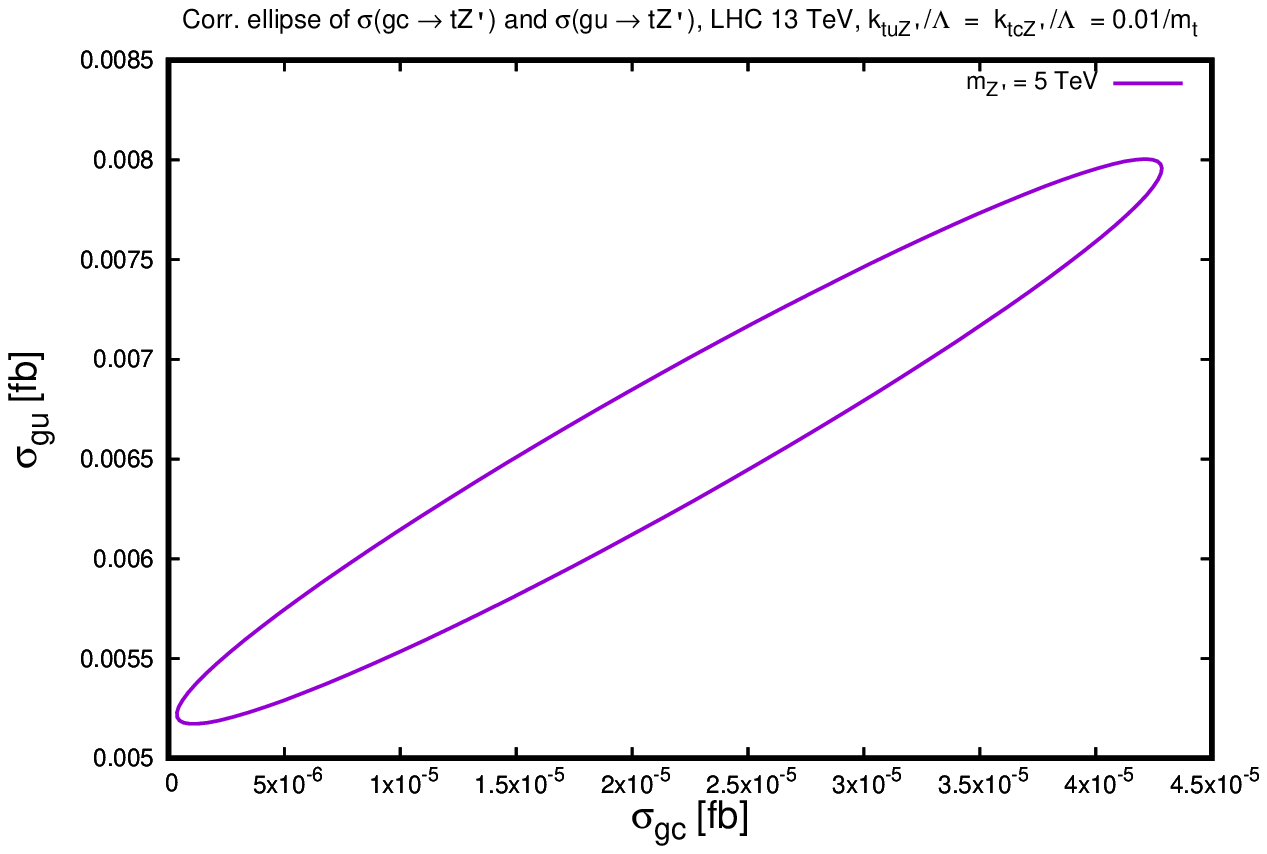}
\hspace{1mm}
\includegraphics[width=8.3cm]{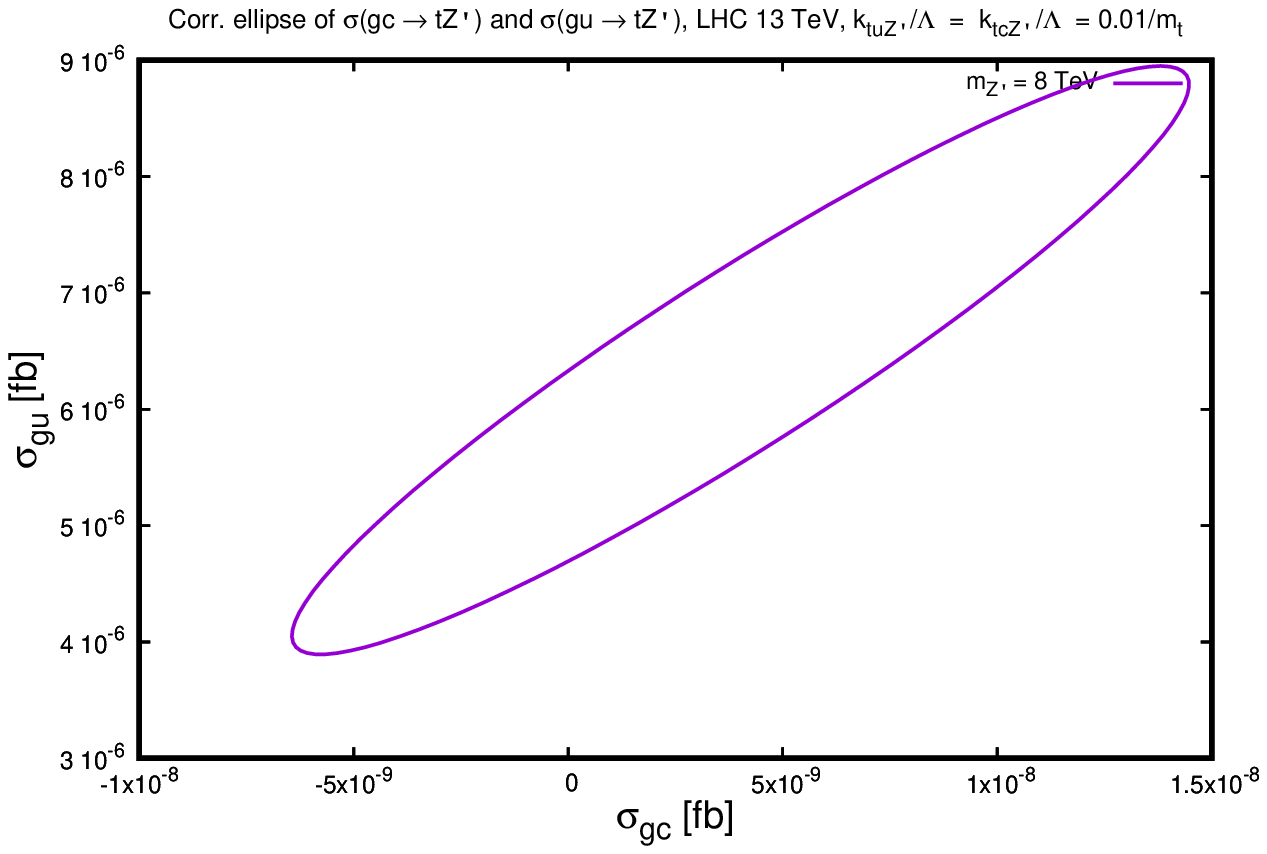}
\caption{Correlation ellipses of the $gu\rightarrow tZ'$ and $gc\rightarrow tZ'$ channels at $\sqrt{S}=13$ TeV.
The figures show CT14NNLO PDF induced uncertainty boundaries of the aNNLO results. Uncertainties are rescaled at
the 68\% C.L. for different values of the $Z^\prime$ mass.}
\label{ellipse-sgu-sgc}
\end{center}
\end{figure}

\begin{figure}[tbh!]
\begin{center}
\includegraphics[width=8.3cm]{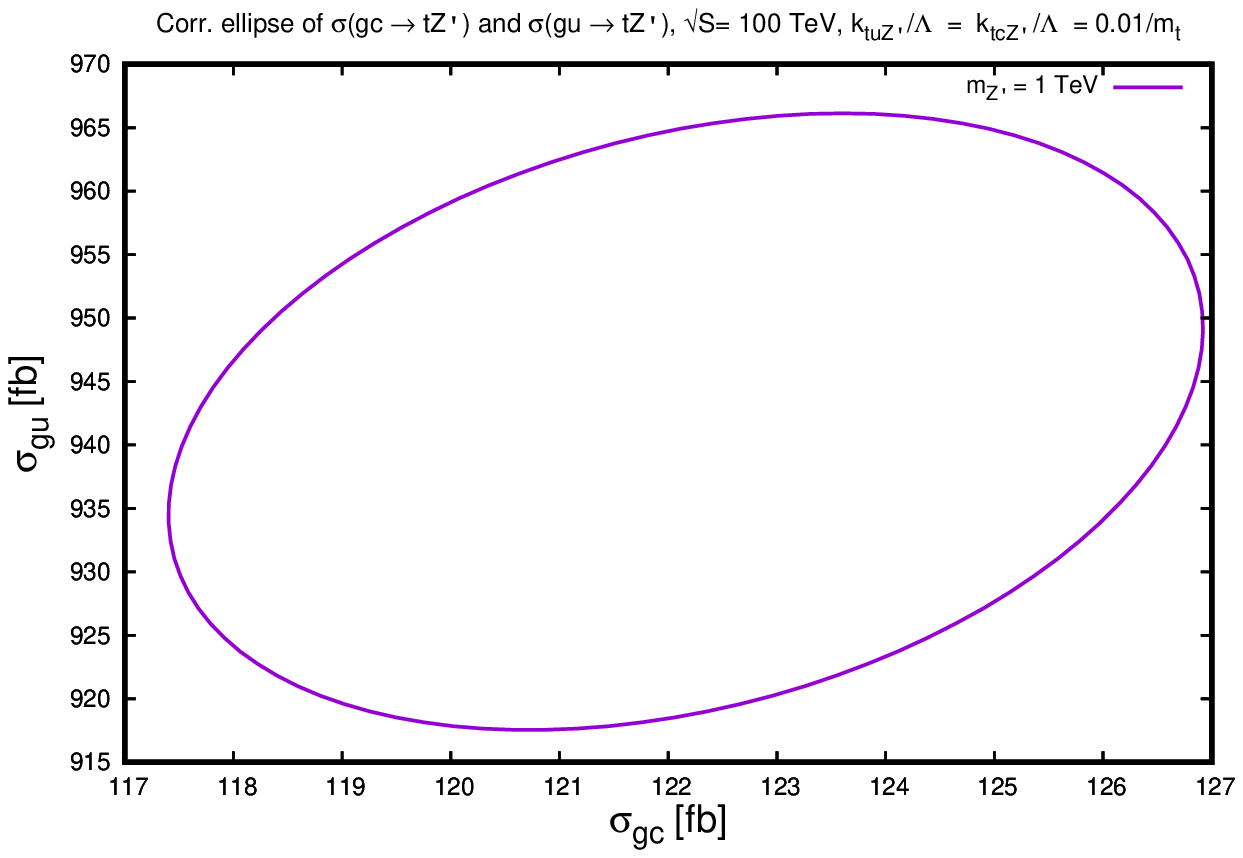}
\hspace{1mm}
\includegraphics[width=8.3cm]{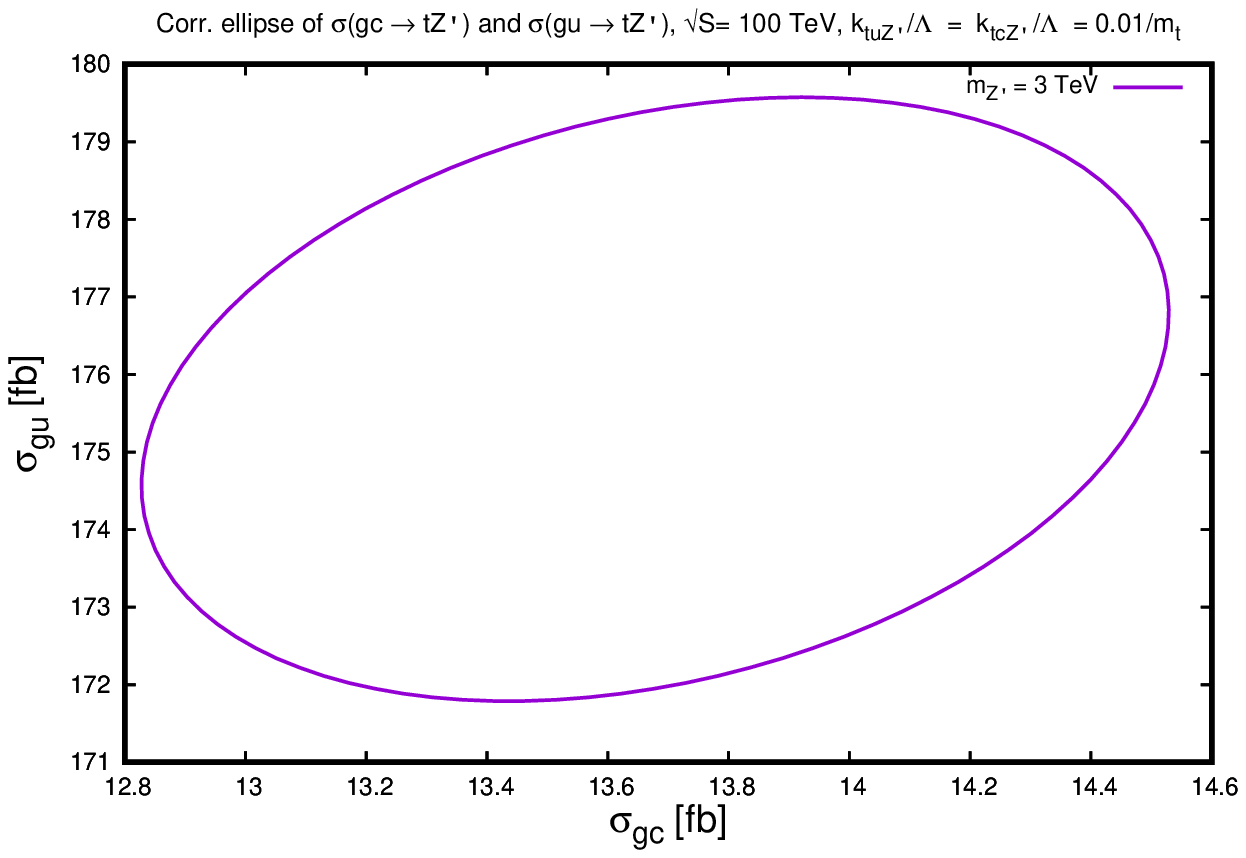}\\
\includegraphics[width=8.3cm]{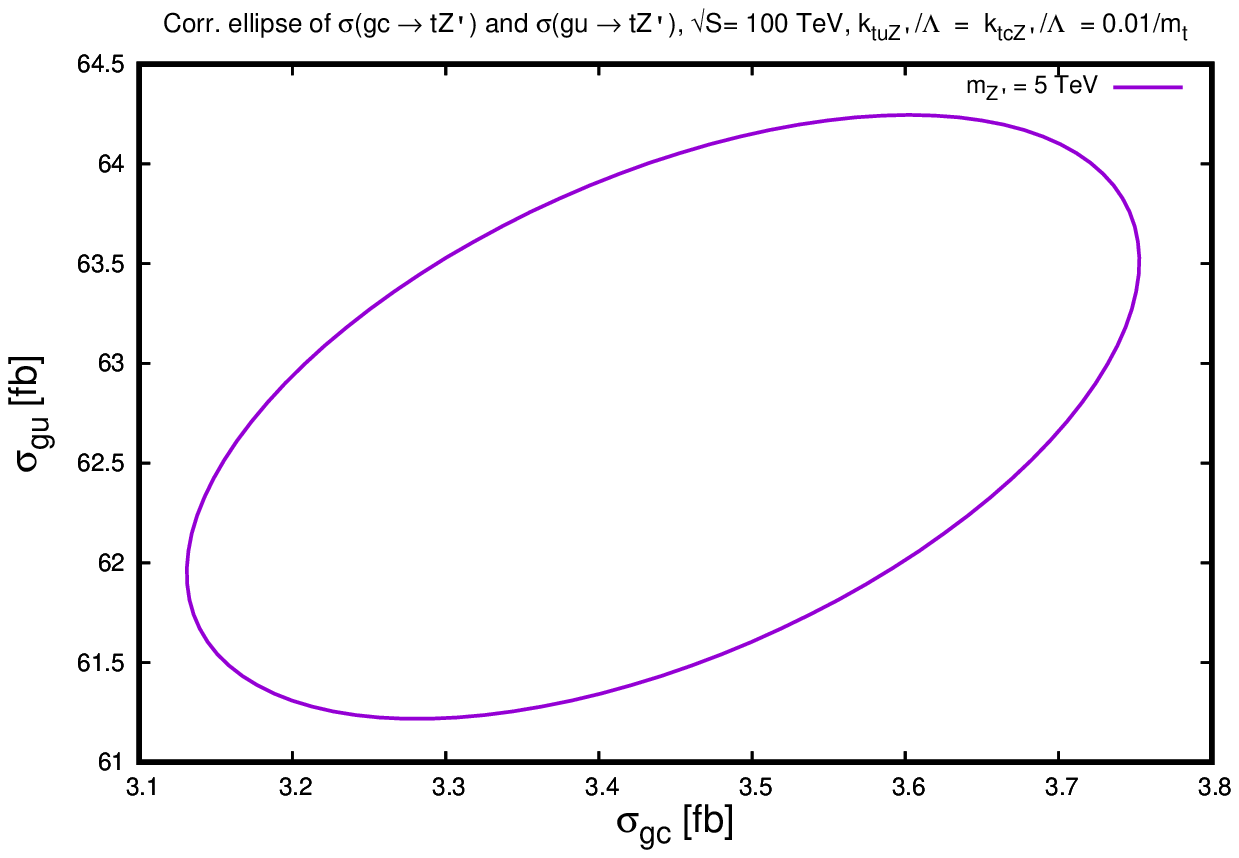}
\hspace{1mm}
\includegraphics[width=8.3cm]{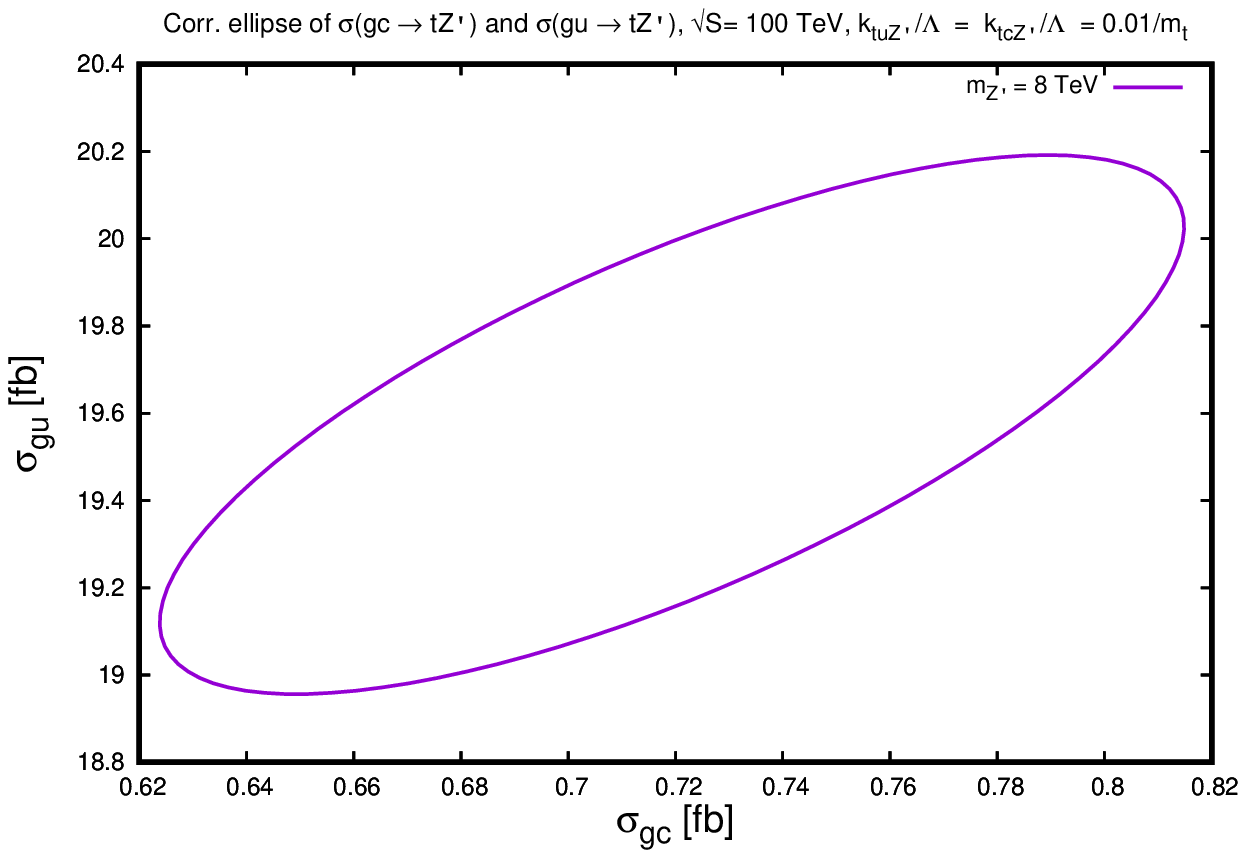}
\caption{Same as in Fig.~\ref{ellipse-sgu-sgc}, but for $pp$ collisions at $\sqrt{S}=100$ TeV.}
\label{ellipse-sgu-sgc-100TeV}
\end{center}
\end{figure}

Besides the correlation with PDFs, important information can also be gathered
from the study of simultaneous uncertainty boundaries of the cross section of
the $gu$ and $gc$ channels. The allowed regions are represented by correlation
ellipses which can be compared to pseudo data in BSM simulations and explore the implications of the PDFs for this process.
In Fig.~\ref{ellipse-sgu-sgc} and \ref{ellipse-sgu-sgc-100TeV} we show the elliptical confidence regions, at 68\% CL,
in $pp$ collisions at 13 and 100 TeV, for $m_{Z'}=1$, 3, 5, and 8 TeV.
These can be used to read off PDF uncertainties and correlations for each pair of cross sections.
At $\sqrt{S}=$ 13 TeV, we notice that the two channels are highly correlated and the induced PDF
uncertainties on the $\sigma_{gc}$ channel are very large for this choice of the collider energy.
This is reflected by the fact that there is a small portion of the ellipse where the PDF induced
errors on the cross sections are larger than the cross section central value itself, allowing for negative values.
At $\sqrt{S}=$ 100 TeV, the $gu$ and $gc$ channels are still highly correlated, but the
induced PDF uncertainties on both the cross sections are smaller as in this kinematic
domain the PDFs are probed at intermediate $x$ where they are better constrained.

\begin{figure}[tbh!]
\begin{center}
\includegraphics[width=8.2cm]{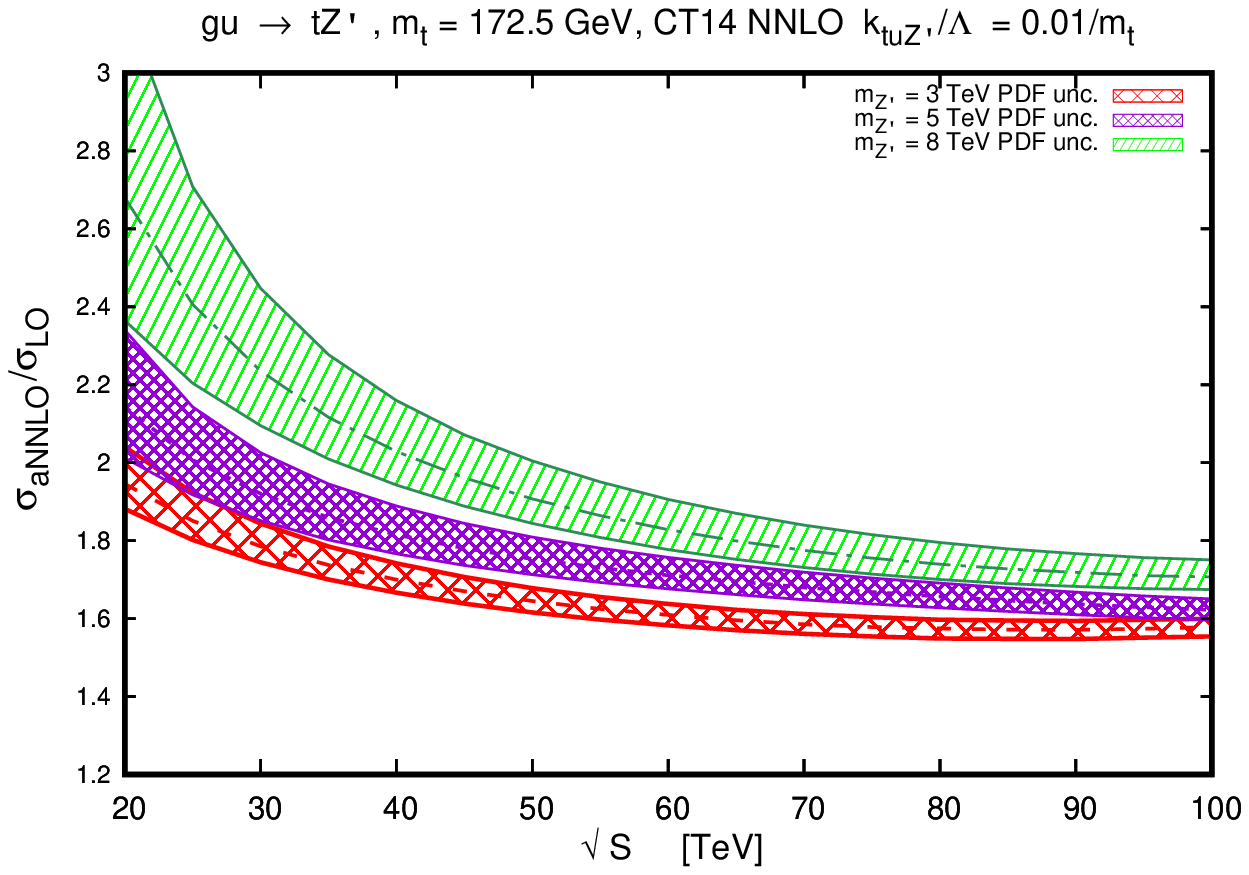}
\includegraphics[width=8.2cm]{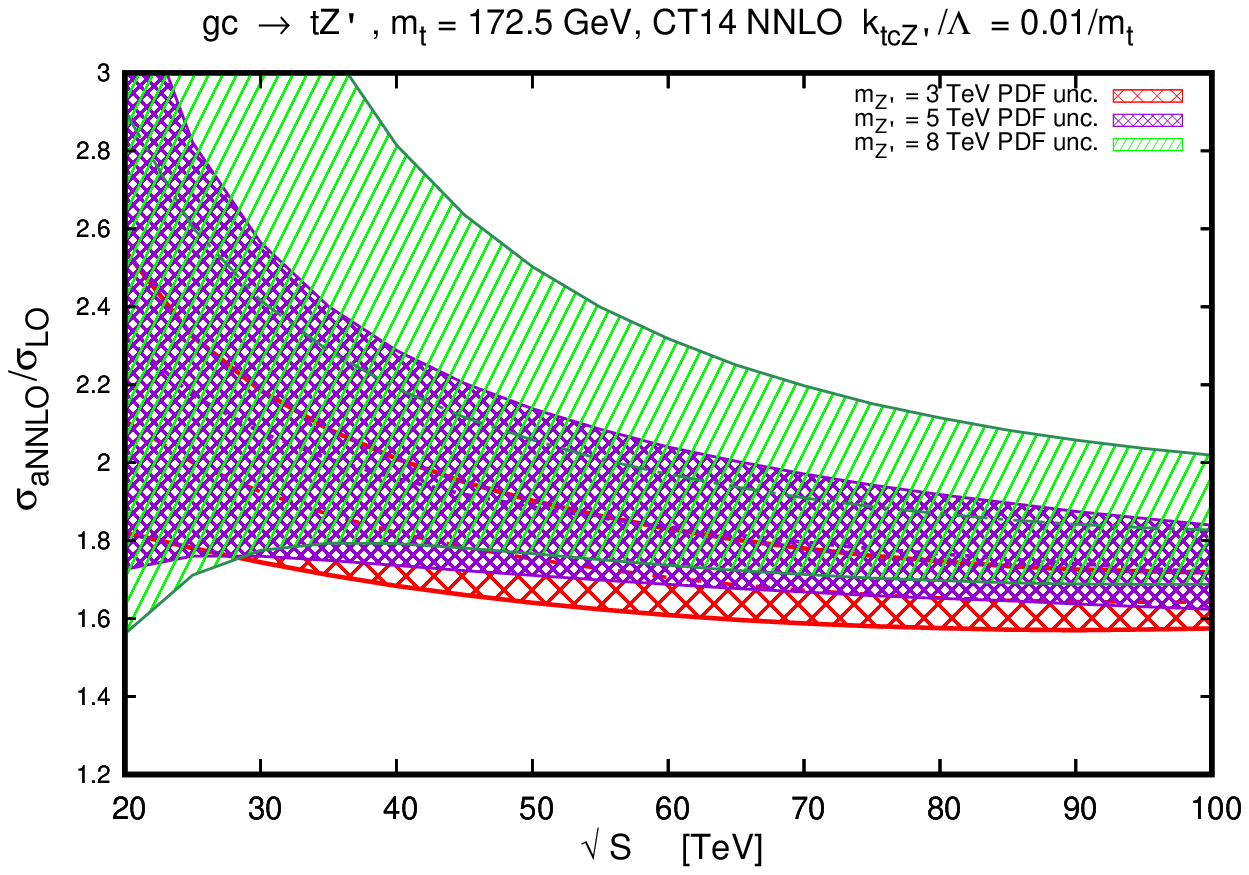}
\caption{$K$-factors with CT14NNLO PDF uncertainties (68\% C.L.) for the $gu\rightarrow tZ'$ and $gc\rightarrow tZ'$ channels.
The figures show a scan in the center of mass energy of the collisions $\sqrt{S}$ for different values of the $Z^\prime$ mass.  }
\label{K-fact-PDFunc}
\end{center}
\end{figure}

\begin{figure}[th!]
\begin{center}
\includegraphics[width=8.3cm]{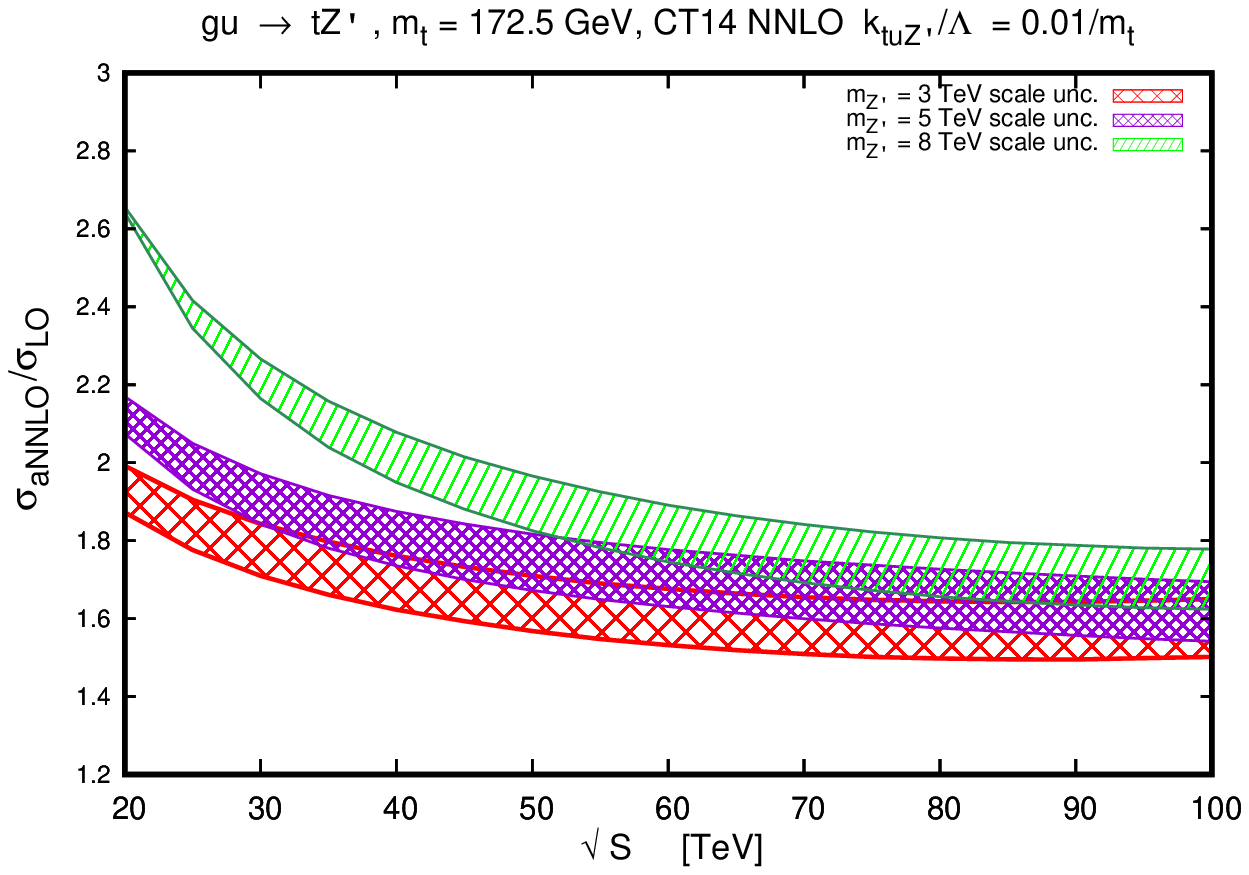}
\includegraphics[width=8.3cm]{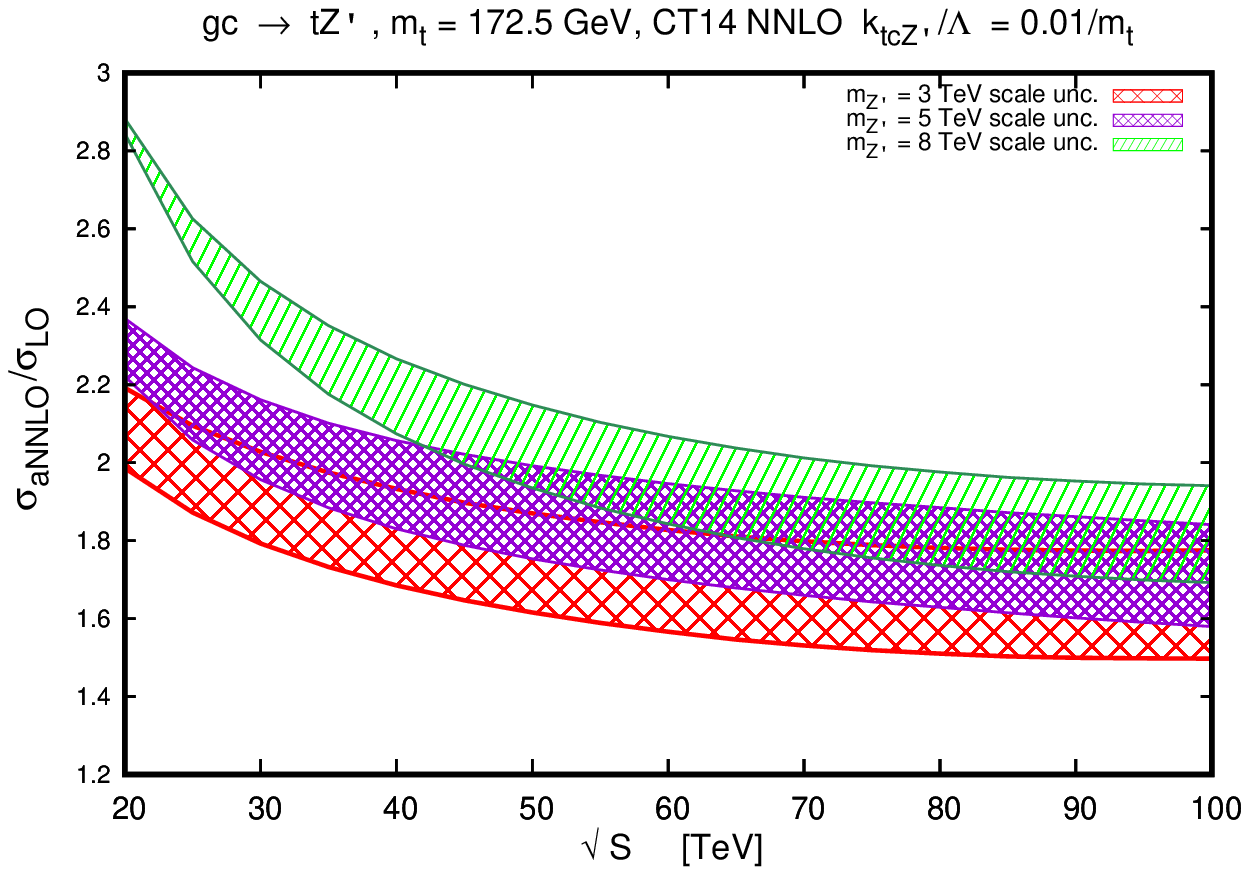}
\caption{Scale uncertainty in the aNNLO $K$-factors for the $gu\rightarrow tZ'$ and $gc\rightarrow tZ'$ channels.
The figures show a scan in the center of mass energy of the collisions $\sqrt{S}$ for different values of the $Z^\prime$ mass.
Scale variation refers to $m_{Z'}/2\leq \mu \leq 2 m_{Z'}$. CT14NNLO PDFs are used.}
\label{K-fact-scaleunc}
\end{center}
\end{figure}

Next, we study the impact of the scale and PDF uncertainties on the aNNLO/LO
$K$-factors as functions of the collider energy for large $\sqrt{S}$ values and
different values of $m_{Z'}$. In Fig.~\ref{K-fact-PDFunc} and \ref{K-fact-scaleunc},
we illustrate the $K$-factors for the $gu$ and $gc$ channels with CT14NNLO PDF and scale
uncertainties respectively. Scale variation refers to $m_{Z'}/2\leq \mu \leq 2 m_{Z'}$ as before.
In Fig.~\ref{K-fact-PDFunc} the PDF uncertainties for each $m_{Z'}$ value are shown using bands
with different hatches and color. At collider energies
below 20 TeV PDF uncertainties are large because PDFs are probed in the large-$x$ region.
In the $gc$ channel, PDF uncertainties are dominant because the charm-quark PDF is less
constrained with respect to the gluon and $u$-quark. In Fig.~\ref{K-fact-scaleunc} the
scale dependence in the aNNLO $K$-factors for the $gu$ and $gc$ channels is illustrated separately.

\clearpage

\subsection{\label{Stringy-Zprimes} String-inspired $Z'$s: $gt \rightarrow tZ'$}

In this section we discuss the phenomenological results obtained from the study of $tZ'$ production where
the $Z'$ originates from a low-energy realization of string-inspired models.
The interaction Lagrangian in Sec.~\ref{stringyZprime}, and the choice of the parameters we have examined,
are based on the models published in Refs.~\cite{Faraggi:2015iaa,Coriano:2008wf}.
These models have not been searched for by the ATLAS and CMS collaborations to the best of our knowledge, therefore the current limits on
the $Z'$ mass and couplings should in principle not be applied here.
The $Z'$ models described in Refs.~\cite{Faraggi:2015iaa,Coriano:2008wf} allow for non-sequential solutions (i.e. charge assignments which are not
proportional to the hypercharge) that are phenomenologically interesting and could in principle be considered in future analyses by both ATLAS and CMS.

An accurate determination of the $gt \rightarrow tZ'$ cross section can play an important role to set constraints on the couplings of $Z'$ to the fermion sector.
In fact, this process can in principle be used together with $Z'$ production in Drell-Yan to
remove the degeneracy between quark and lepton couplings~\cite{Petriello:2008zr,Petriello:2008pu}.

The leading-order cross section is given by the $s$- and $t$-channels of the $gt \rightarrow tZ'$ process
and the structure of the couplings is given in Sec.~\ref{stringyZprime}.
The $gt \rightarrow tZ'$ process with $m_{Z'}$ in the TeV range requires the top-quark PDF in the initial state.
In our phenomenological application, $\mu=m_{Z'}\gg m_t$ and we consider the top quark as an active flavor inside the proton with very good approximation.
Therefore, in the rest of this analysis we work with the $6$-flavor scheme and use the
NNPDF3.1 PDFs~\cite{Ball:2017nwa} with $n_f=6$ and $\alpha_s(m_Z)=0.118$, where $n_f$ is the number of active flavors.
We set $m_t=0$ in the initial state lines in the calculation of the LO cross section.
In this case, PDF uncertainties are calculated at 1-$\sigma$ C.L. (see Appendix~\ref{AppA}) which is almost identical
to the 68\% C.L. in absence of statistical fluctuations in the determination of the PDFs.

\begin{figure}[htb!]
\begin{center}
\includegraphics[width=8cm]{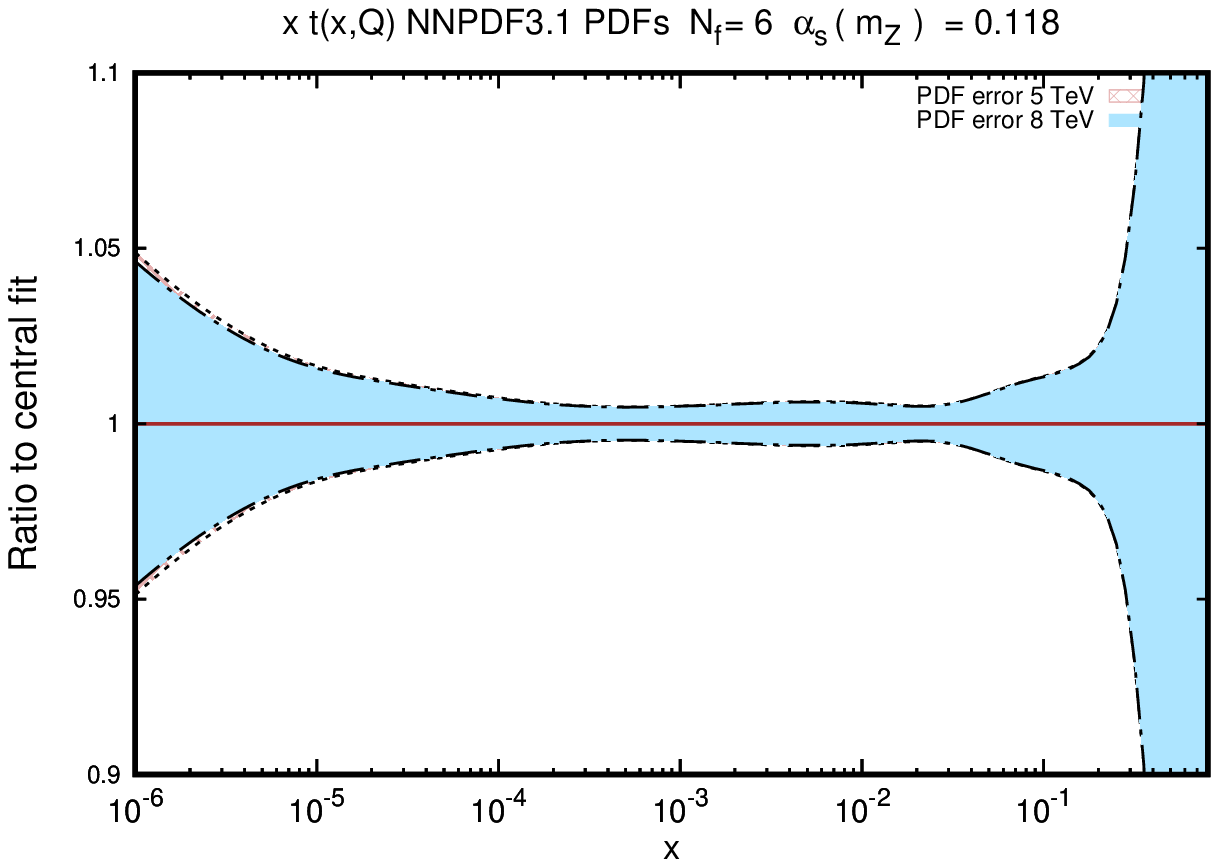}
\includegraphics[width=8cm]{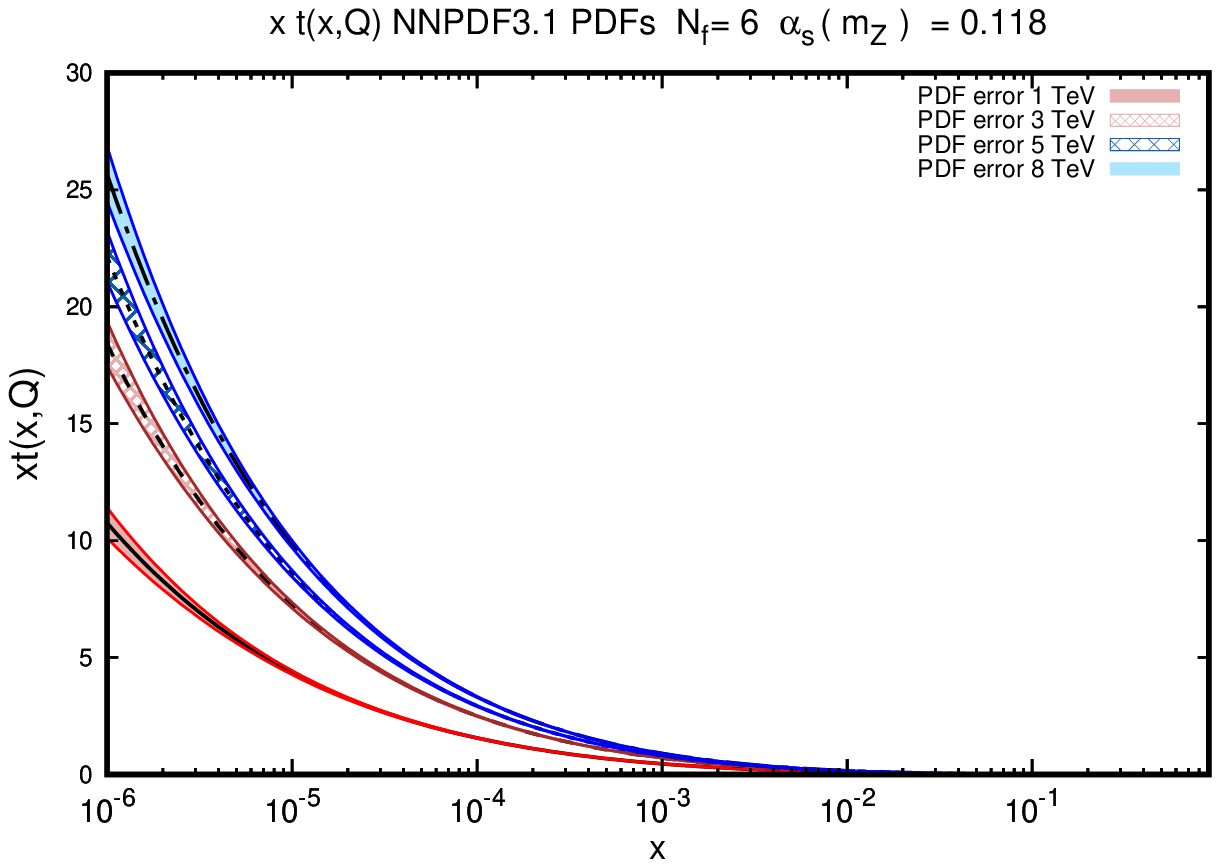}
 \caption{The error bands represent NNPDF3.1 NNLO $n_f=6$ PDF uncertainties evaluated at the 1-$\sigma$ C.L.. }
\label{top-quark-PDF}
\end{center}
\end{figure}

In Fig.~\ref{top-quark-PDF} we illustrate the top-quark PDF uncertainty as a function of $x$ for different values of the final-state $Z'$ mass.
The $gt \rightarrow tZ'$ process probes the top-quark and gluon PDFs at large $x$ where uncertainties are large at the LHC Run II collision energies.
Precision measurements in the extended kinematic domain of the future FCC-eh collider will allow us to extract PDFs at large $x$
for the individual quark flavors at the percent level precision. The precision of the top-quark PDF will be improved in this kinematic region
enhancing the FCC-hh discovery potential of $Z'$s with mass of ${\cal O}$(10) TeV also in rare processes.

\begin{figure}
\begin{center}
\includegraphics[width=8cm]{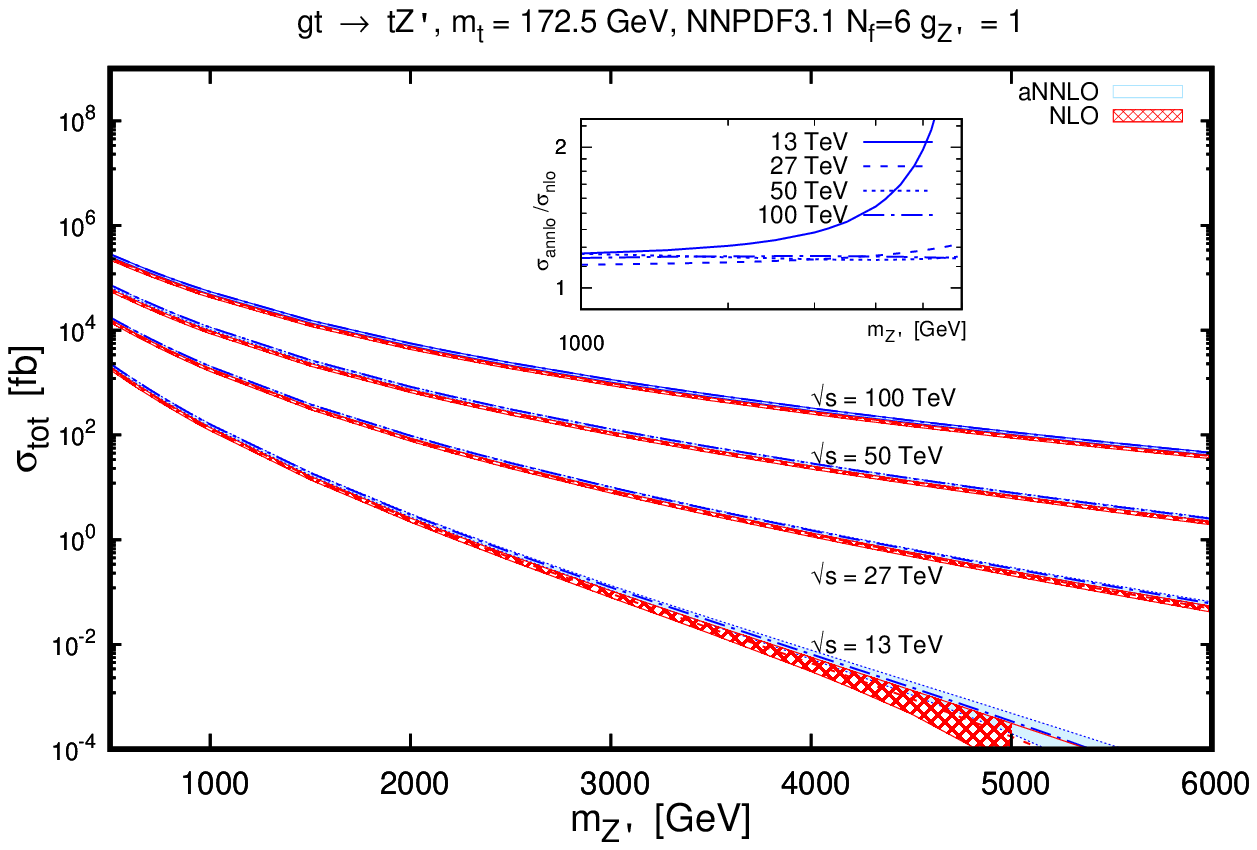}
\includegraphics[width=8cm]{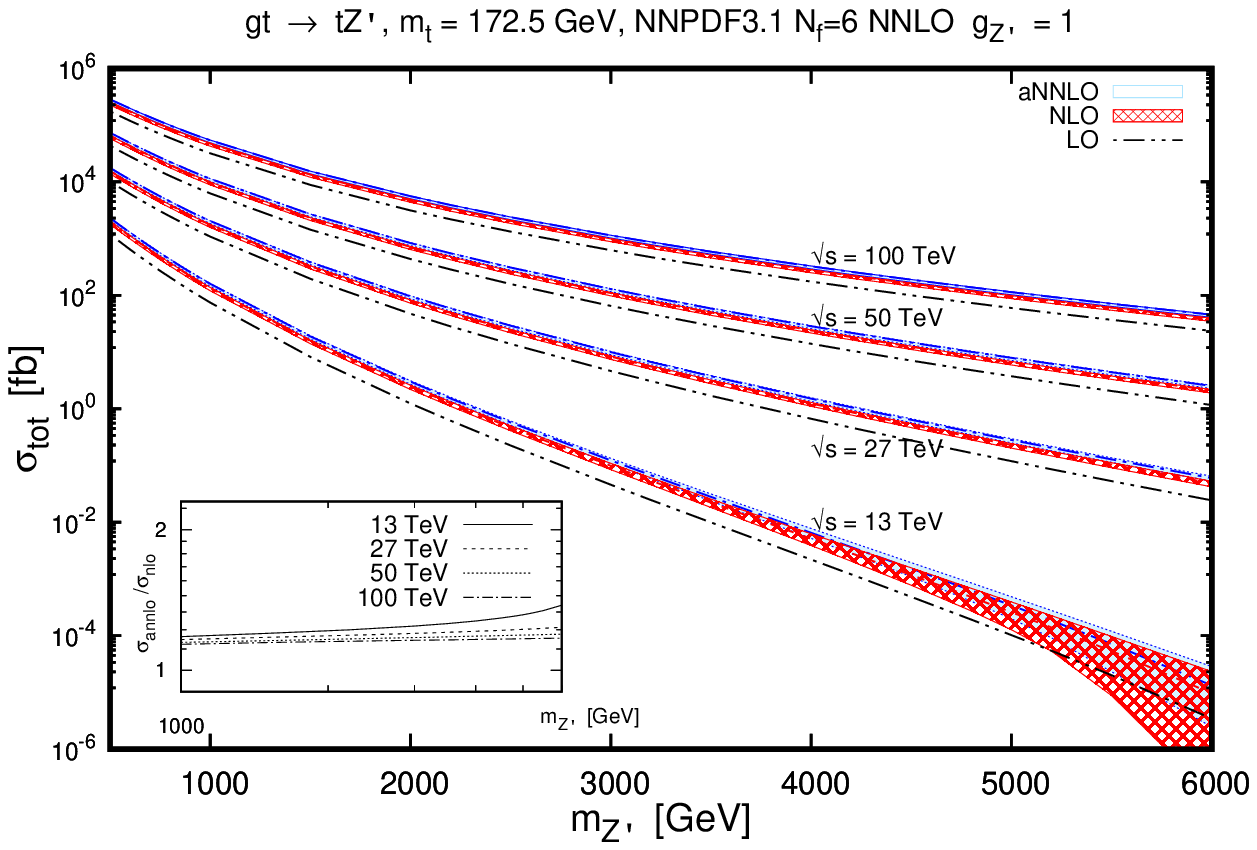}
\caption{(Left) Total cross sections for the $gt\rightarrow tZ'$ process at NLO and aNNLO
as a function of the mass of the $Z'$ for various collider energies.
The aNNLO result is obtained using NNPDF3.1 NNLO $n_f=6$ PDFs, while NLO is obtained using the same PDFs at NLO.
(Right)  Total cross sections for the $gt\rightarrow tZ'$ process at LO, NLO, and aNNLO
where all results use NNPDF3.1 NNLO PDFs.
In both plots the error bands represent PDF uncertainties at 1-$\sigma$ C.L.,
and the inset plots show $\sigma_{aNNLO}/\sigma_{NLO}$ $K$-factors.}
\label{tZp-mazscan-nnlopdfs-vs-nlopdfs}
\end{center}
\end{figure}

The left plot of Fig.~\ref{tZp-mazscan-nnlopdfs-vs-nlopdfs} shows the NLO and aNNLO total cross section
for the $gt\rightarrow tZ'$ process as a function of $m_{Z'}$ at collider energies $\sqrt{S} = 13, 27, 50, 100$ TeV.
The error bands represent the induced PDF uncertainties on the cross section at 1-$\sigma$ C.L.
obtained by using NNPDF3.1 $n_f = 6$ PDFs. The aNNLO prediction is obtained using NNPDF3.1 NNLO $n_f=6$ PDFs,
while the NLO is obtained using NNPDF3.1 NLO $n_f=6$ PDFs. The LO cross section is not shown here because the NNPDF3.1 $n_f=6$ PDFs at LO
are not available. The inset plot shows the $\sigma_{aNNLO}/\sigma_{NLO}$ $K$-factors from where we observe that the $K$-factors are large and they increase as $m_{Z'}$ increases, and they decrease when the collider energy increases, as for the case of the FCNC $Z'$s.

In the plot on the right of Fig.~\ref{tZp-mazscan-nnlopdfs-vs-nlopdfs}, the cross section is obtained by convoluting hard scatterings at LO, NLO, and aNNLO,
with NNLO PDFs in order to show the enhancement due to the hard-scattering ontributions only.

\begin{figure}
\begin{center}
\includegraphics[width=8.2cm]{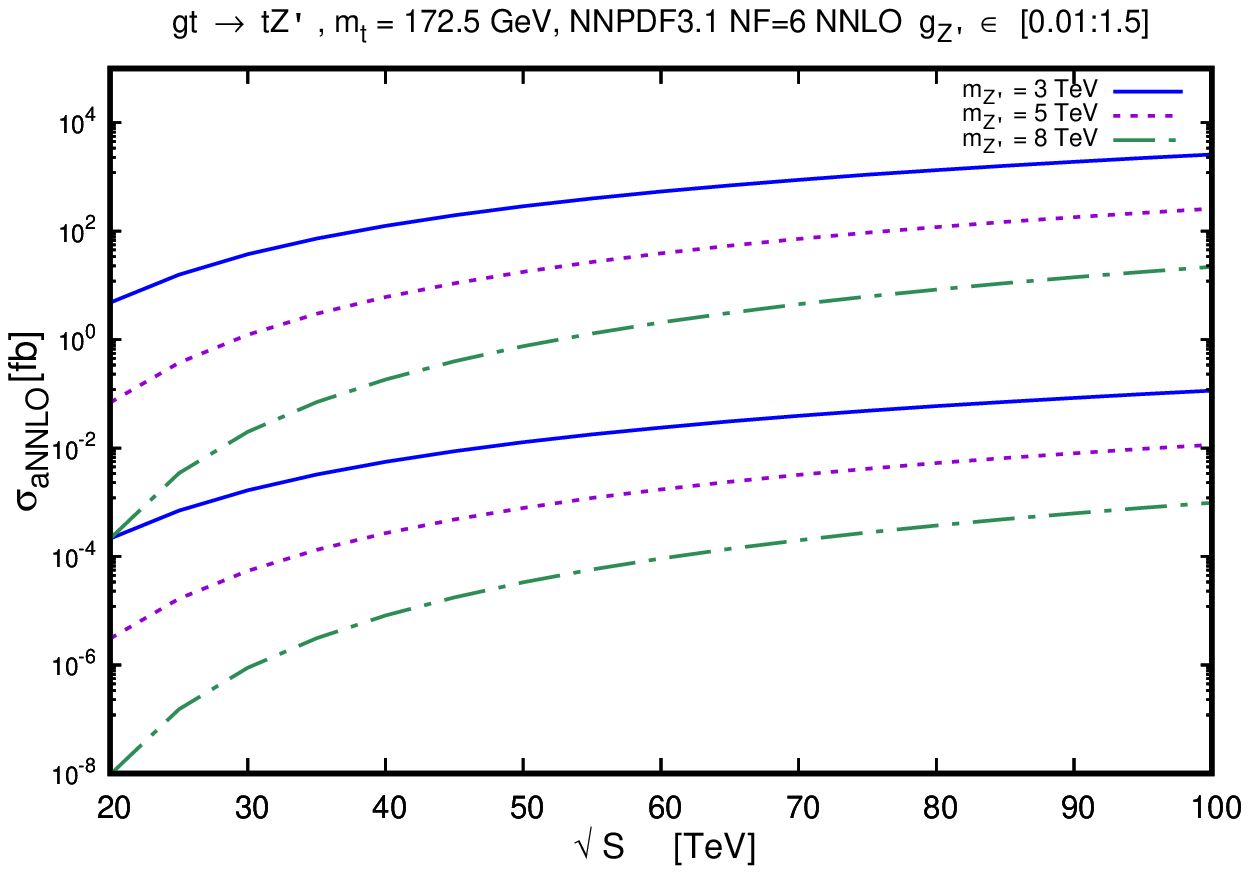}
\includegraphics[width=8.2cm]{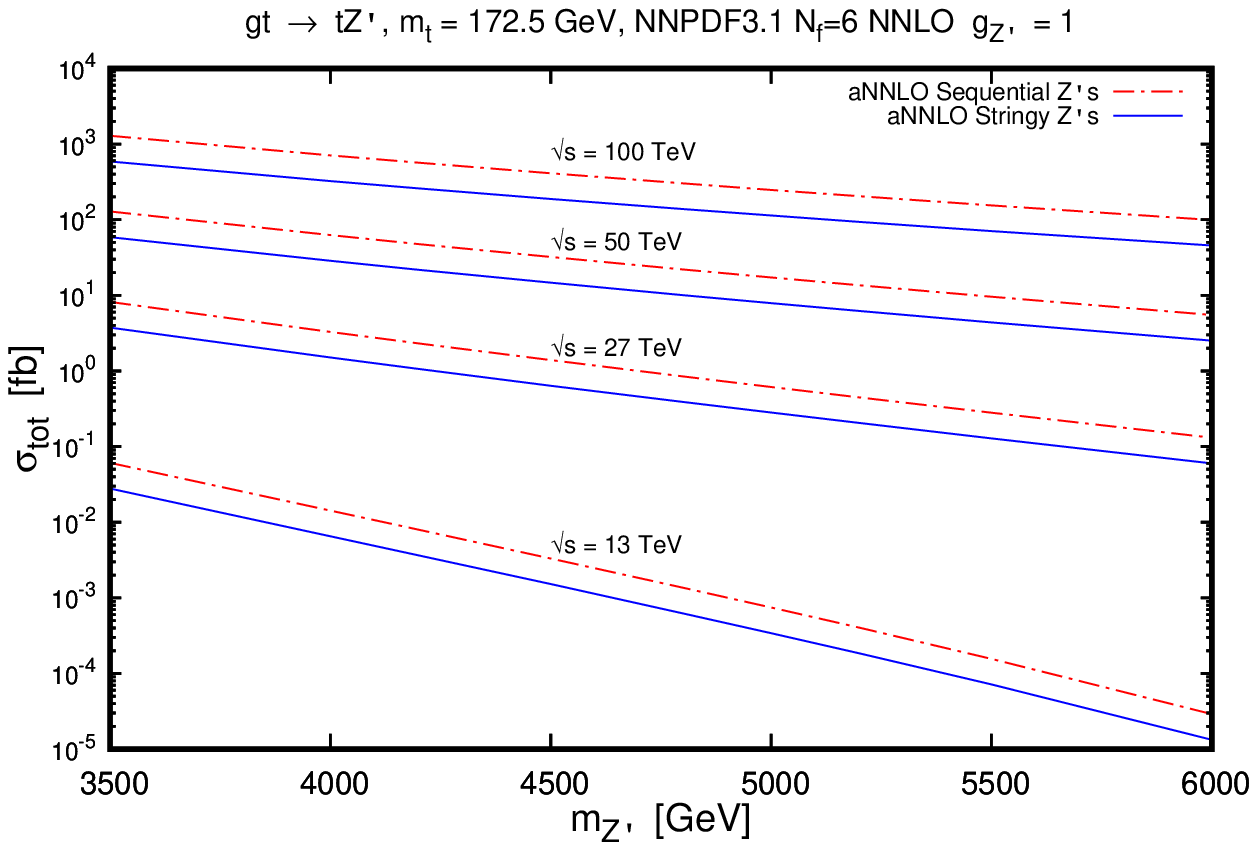}
\caption{Left: Scan of the $g_{Z'}$ parameter for the $gt\rightarrow tZ'$ process.
The plot shows results of the total cross section at aNNLO as a function of collider energy.
Bands with different dashing represent $Z'$ mass values of 3, 5, and 8 TeV.
Right: Comparison between string-inspired $Z'$s and sequential
$Z'$s for different values of the collider energy at aNNLO.}
\label{gz-scan-tZp-mazscan}
\end{center}
\end{figure}

The $Z'$ coupling $g_{Z'}$ is considered as a free parameter and as a case study we choose
$g_{Z'}=1$ as the default choice. A $g_{Z'}$ parameter scan is illustrated in Fig.~\ref{gz-scan-tZp-mazscan} (left)
where the aNNLO cross section is plotted as a function of $\sqrt{S}$ for different values of $m_{Z'}$ which
correspond to bands with different dashing. We explore $g_{Z'}$ variations in $0.01\leq g_{Z'}\leq 1.5$ and
observe that when $g_{Z'}$ varies the cross section is basically rescaled and it spans approximately two orders of magnitude.

Moreover, for comparison purposes, we consider the production of a sequential $Z'$ as a commonly-used point of reference.
In Fig.~\ref{gz-scan-tZp-mazscan} (right) we illustrate a comparison between aNNLO total cross sections for the production of string-inspired $Z'$s and the production of sequential $Z'$s, for different values of the collider energy. The sequential $Z'$s are extra neutral vector bosons
which have vector and axial-vector couplings equal to those of the SM $Z$-boson, but such that
their right-handed and left-handed couplings to quarks are defined up to a constant factor
which we set equal to $g_{Z'}$, e.g., $g^{(Z')}_{R,L} = g_{Z'}(g_V\pm g_A)$.
In this specific comparison we consider $Z'$ masses larger than 4 TeV because sequential
$Z'$s are currently excluded for smaller masses~\cite{Aaboud:2017buh,CMS:2016abv}. As expected, the shapes in the two models are identical.

\begin{figure}
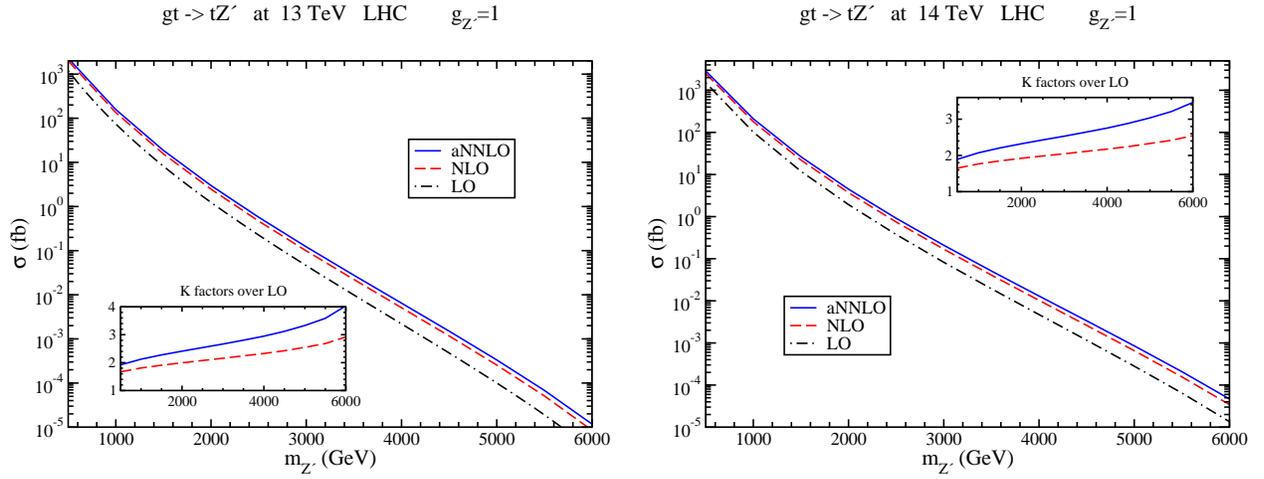

\begin{center}
\includegraphics[width=8cm]{tZp13lhcnewplot.eps}
\hspace{2mm}
\includegraphics[width=8cm]{tZp14lhcnewplot.eps}
\caption{Total cross section for the $gt\rightarrow tZ'$ process at the LHC 13 TeV (left) and 14 TeV (right).
LO, NLO and NNLO calculations are obtained using NNPDF3.1 NNLO $n_f=6$ PDFs.}
\label{tZp13-14-Kfac}
\end{center}
\end{figure}

\begin{figure}
\begin{center}
\includegraphics[width=10cm]{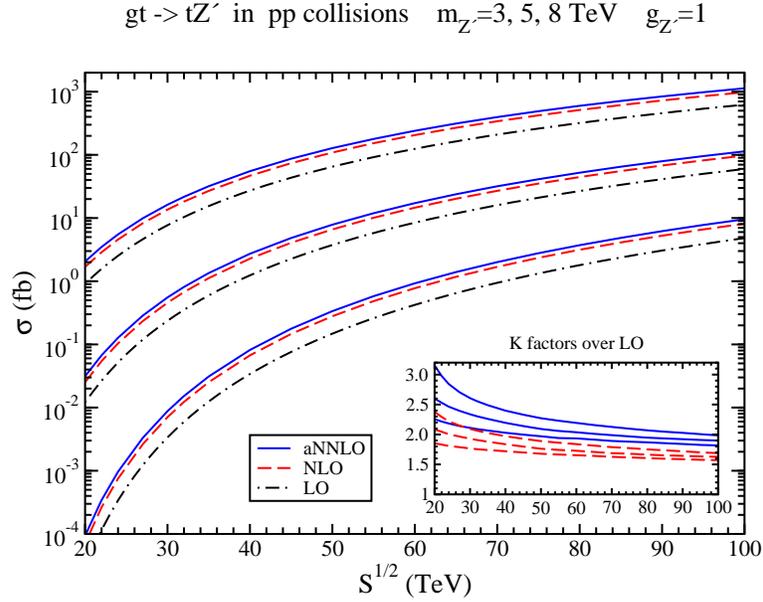}
\caption{Total cross sections for the $gt\rightarrow tZ'$ process. The plot shows results
as a function of collider energy for three choices of $Z'$ mass, 3, 5, and 8 TeV.
The inset plot displays $K$-factors. NNPDF3.1 NNLO $n_f=6$ PDFs are used for LO, NLO, and aNNLO calculations.
Cross sections get smaller with increasing $Z'$ mass while $K$-factors get larger.}
\label{tZprootS}
\end{center}
\end{figure}

Prospects at the LHC at 13 and 14 TeV collision energies are shown in Fig.~\ref{tZp13-14-Kfac} where the inset plots show
the NLO/LO and aNNLO/LO $K$-factors. Here, LO, NLO, and aNNLO cross sections are all obtained by using NNPDF3.1 NNLO $n_f=6$
PDFs to show the soft-gluon enhancement in the hard scattering.
We note the large effect of the higher-order corrections, which more than triple the LO result for a 6 TeV $Z'$ mass.
We also provide numerical values for the $gt\rightarrow tZ'$ cross section and $K$-factors at 13 TeV energy in Table \ref{tZ'table-gt} of Appendix~\ref{AppC}.
Total cross section results as functions of the collider energy up to 100 TeV for different
values of $m_{Z'}$ are given in Fig.~\ref{tZprootS}.
The inset plot shows the NLO/LO and aNNLO/LO $K$-factors where NNPDF3.1 NNLO $n_f=6$ PDFs are used for LO, NLO, and aNNLO calculations.
While the cross sections get smaller with increasing $Z'$ mass,
the $K$-factors get larger because this kinematic region is closer to the partonic threshold.

\begin{figure}
\begin{center}
\includegraphics[width=8.3cm]{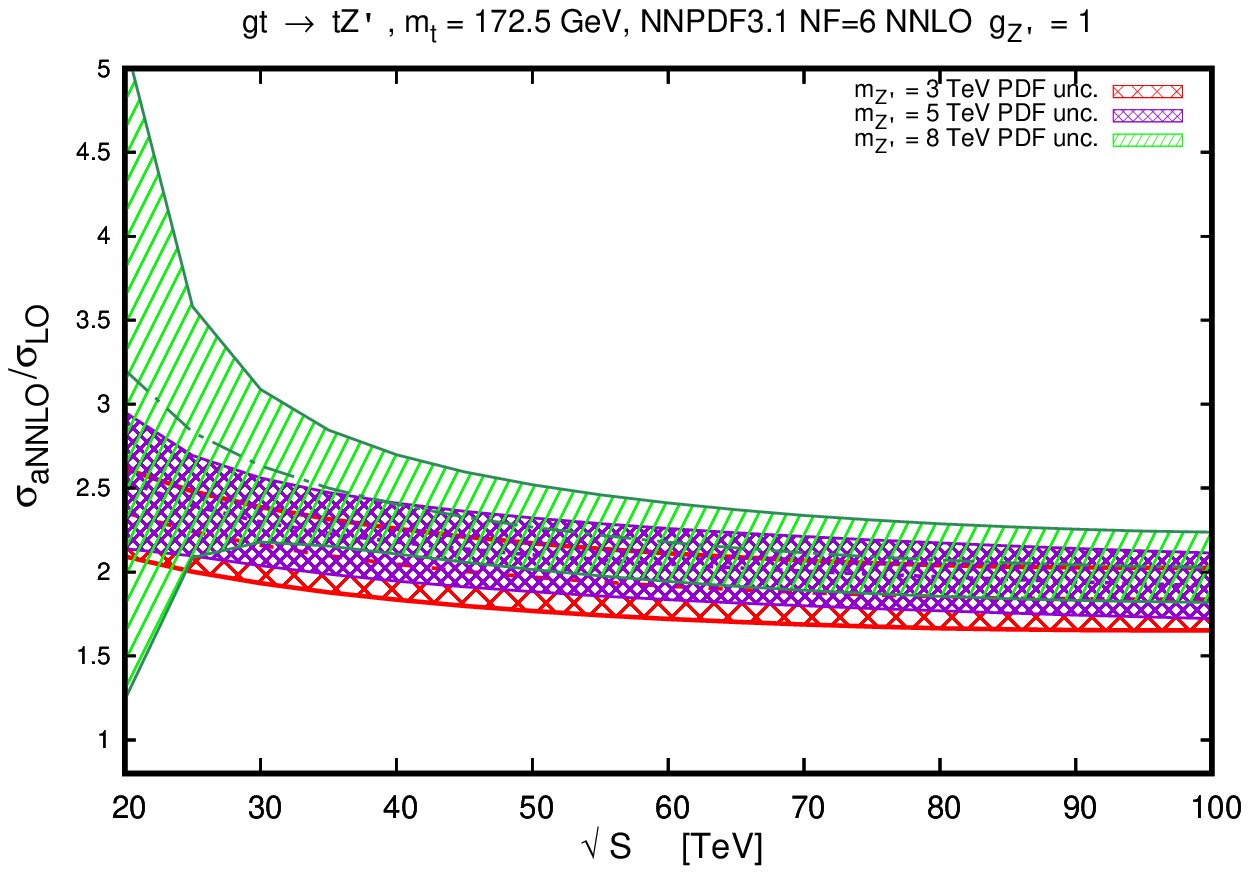}
\includegraphics[width=8.3cm]{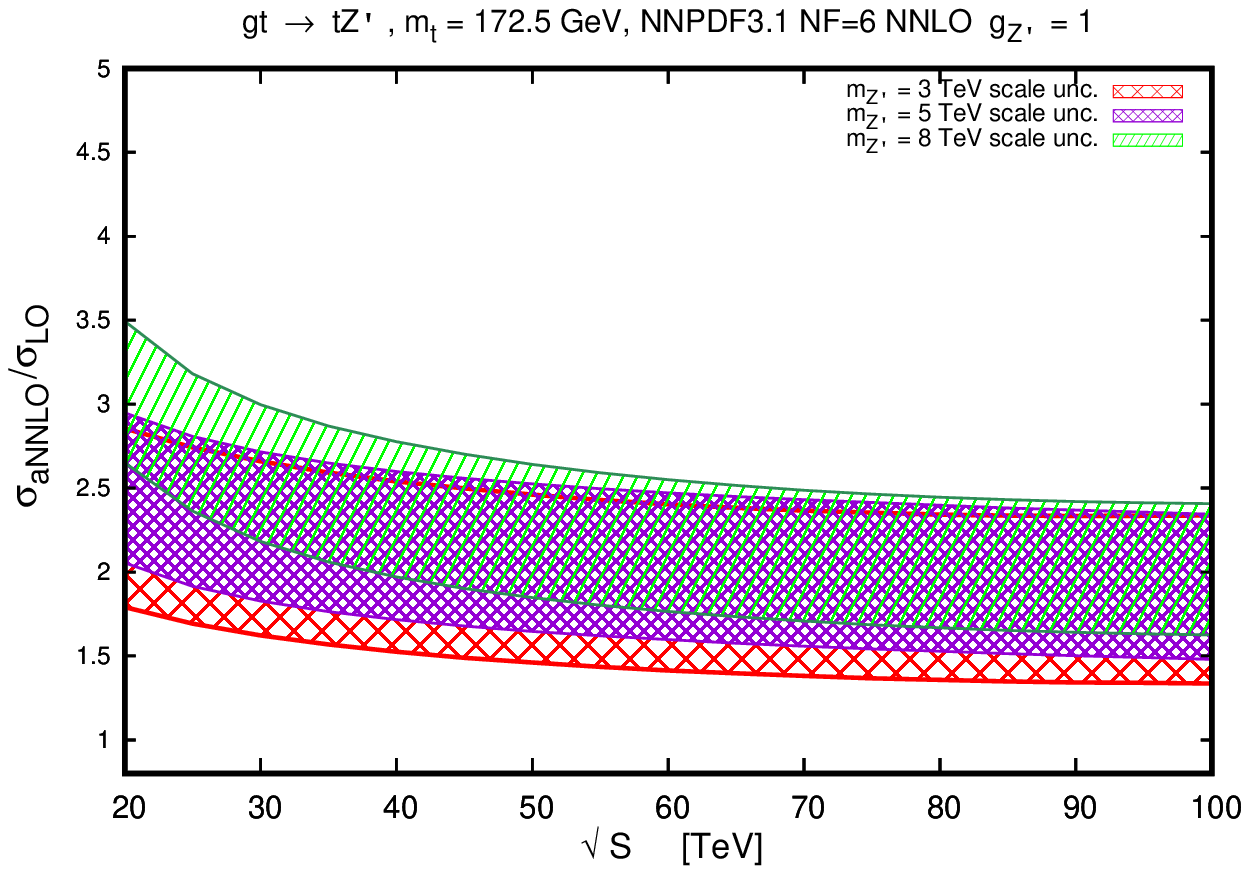}
\caption{Left: $K$-factors for the $gt\rightarrow tZ'$ process with NNPDF3.1 $n_f=6$ PDF uncertainties.
The plot shows aNNLO/LO results as a function of $\sqrt{S}$ for three choices of $Z'$ mass, 3, 5, and 8 TeV.
Right: $K$-factors with scale uncertainty bands. Scale variation refers to $m_{Z'}/2\leq \mu \leq 2 m_{Z'}$. }
\label{tZprootS-PDFunc-scaleunc}
\end{center}
\end{figure}

In the left plot of Fig.~\ref{tZprootS-PDFunc-scaleunc} we illustrate the induced NNPDF3.1 $n_f=6$ NNLO PDF uncertainty on
the $\sigma_{\textrm{aNNLO}}$ total cross section which we normalize to $\sigma_{\textrm{LO}}$ to obtain $K$-factors.
Here the LO cross section is also obtained with NNLO PDFs.
The large uncertainty of the top-quark NNLO PDF dominates at all collider energies and for every value of $m_{Z'}$.
In the plot on the right of Fig.~\ref{tZprootS-PDFunc-scaleunc} we show the scale uncertainty due to factorization scale variation in $m_{Z'}/2\leq \mu \leq 2 m_{Z'}$.
As mentioned in previous sections, the $K$-factors here are defined as $\sigma_{\textrm{aNNLO}}(\mu)/\sigma_{\textrm{LO}}$ where $\sigma_{\textrm{LO}}$
is obtained using the default central choice $\mu=m_{Z'}$ and NNPDF3.1 $n_f=6$ NNLO PDFs.

\subsubsection{Top-quark $p_T$ distributions for string-inspired $Z'$s}

\begin{figure}
\begin{center}
\includegraphics[width=10cm]{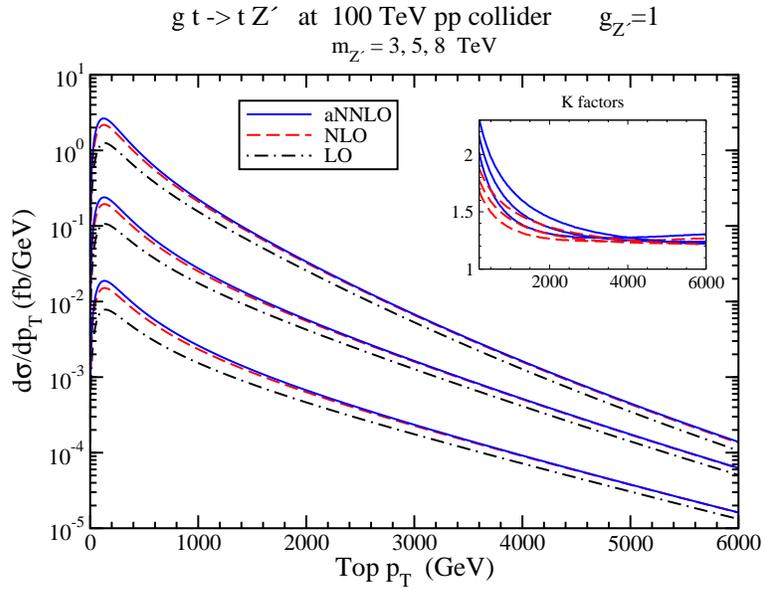}
\caption{Top-quark $p_T$ distributions for the $gt\rightarrow tZ'$ process at 100 TeV $pp$ collider energy for $m_{Z'}=3$, 5, and 8 TeV.
NNPDF3.1 NNLO $n_f=6$ PDFs are used for LO, NLO, and aNNLO calculations.}
\label{pttZp100}
\end{center}
\end{figure}

In this section we show the top-quark $p_T$ distributions for this process.
Fig.~\ref{pttZp100} shows the top-quark $p_T$ distributions in the $gt \rightarrow tZ'$
process at LO, NLO, and aNNLO for different $m_{Z'}$ values at a collider energy of 100 TeV.
NNPDF3.1 NNLO $n_f=6$ PDFs are used for LO, NLO, and aNNLO calculations to emphasize the enhancement in the hard scattering contribution.
The $K$-factors are shown in the inset plot.

\clearpage

\mysection{\label{Conclusions} Conclusions}

We have studied $tZ'$ production in various BSM models at hadron colliders.
We performed a phenomenological QCD analysis where we scrutinized $tZ'$ production in the presence of FCNC and in the case in which the extra $Z'$
is generated within a low-energy realization of string theory models.
We have calculated theoretical predictions for cross sections and top-quark $p_T$ distributions that include higher-order soft-gluon corrections.
In particular, theory predictions are obtained at aNNLO in QCD by extending the soft-gluon resummation formalism to the case in which a
top quark is produced in association with a heavy neutral vector boson in $pp$ collisions at energies that relevant for the LHC and for
future new-generation hadron colliders like FCC-hh and SppC. We have found that QCD corrections due to soft-gluon
emissions are considerable and need to be included in precision studies.

We have investigated the impact of uncertainties due to proton PDFs as well as uncertainties due to scale variation.
PDFs uncertainties represent the major source of uncertainty in this analysis. Moreover, we explored the parameter space
for the BSM models we scrutinized by performing parameter scans and studying the sensitivity of the cross section to parameter changes.
We have found that the total $tZ'$ cross section has large sensitivity on the mass of the $Z'$.

These theoretical results will be useful for $tZ'$ production searches at the LHC and future hadron colliders.

\mysection*{Acknowledgements}
We thank Gauthier Durieux and Fabio Maltoni
for correspondence and suggestions about the use of Madgraph5.
The work of M.G. is supported by the National Science Foundation under Grant No. PHY 1820818.
The work of N.K. is supported by the National Science Foundation under Grant No. PHY 1820795.

\appendix

\mysection{\bf Appendix: PDF uncertainties \label{AppA}}

The CT14NNLO PDF uncertainties are determined within the Hessian method at 90\% C.L., and the CT14NNLO eigenvector sets relative to the positive and
negative excursion of the PDF parameters are determined in the QCD global analysis published in Ref.~\cite{Dulat:2015mca}.
The induced PDF errors on the cross section are obtained by using the asymmetric formula ~\cite{Nadolsky:2001yg}
\begin{eqnarray}
 &  & \delta^{+}\sigma=\sqrt{\sum_{i=1}^{N_a}\left[\textrm{max}\left(\sigma_{i}^{(+)}-\sigma_{0},\sigma_{i}^{(-)}-\sigma_{0},0\right)\right]^{2}},\nonumber \\
 &  & \delta^{-}\sigma=\sqrt{\sum_{i=1}^{N_a}\left[\textrm{max}\left(\sigma_0 - \sigma_{i}^{(+)},\sigma_0 - \sigma_{i}^{(-)},0\right)\right]^{2}},
\end{eqnarray}
in terms of $\sigma_{0}$, the cross section obtained with the best-fit (central) PDF value,
and $\sigma^{\pm}_i$, the cross sections for positive and
 negative variations of the PDF parameters along the $i$-th
 eigenvector direction in the $N_a$-dimensional PDF parameter space.
PDF error bands at 68\% C.L. are obtained by the rescaling factor 1.645.

For the NNPDF3.1 NNLO PDF uncertainties, the central value $F_0$ (where $F_0$ can be a cross section or a PDF)
is given by the average and the standard deviation $\delta F$ is taken over
the observable $F$ calculated with each PDF replica set, $S_k$ ($k = 1$,\dots, $N_{rep}$)~\cite{Watt:2011kp,Alekhin:2011sk,Buckley:2014ana}
\begin{eqnarray}
&&F_0 = \langle F\rangle = \frac{1}{N_{rep}} \sum_{k=1}^{N_{rep}} F(S_k)
\nonumber\\
&& \delta F = \sqrt{\frac{1}{N_{rep} -1} \sum_{k=1}^{N_{rep}}\left(F(S_k) - \langle F\rangle\right)^2}
\nonumber\\
&&\hspace{0.5cm}=\sqrt{\frac{N_{rep}}{N_{rep} -1} \left( \langle F^2\rangle  - \langle F\rangle^2 \right)   } \,.
\end{eqnarray}
The NNPDF3.1 set with $N_f=6$ and $\alpha_s(m_Z)=0.118$ which we have used, containes 100 replicas.
The 68 \% C.L. and 1-$\sigma$ PDF uncertainties are very similar in absence of non-gaussian behavior of the probability distribution.

\mysection{\bf Appendix: Correlations \label{AppB}}

If $A(f)$ and $B(f)$ are two quantities that depend on a generic PDF $f$, determined within the Hessian method, the extent
of correlation between $A$ and $B$ can be assessed by calculating the correlation cosine
\ba
\cos{\phi}_{AB} = \frac{1}{4 \Delta{A}\Delta{B}} \sum_{k=1}^n \hat{A}(f_k) \hat{B}(f_k)\,,
\ea
where
\ba
\hat{A}(f_k)=\left[A(f_k^+) - A(f_k^-)\right],~~~~~ \hat{B}(f_k)= \left[B(f_k^+) - B(f_k^-)\right],
\ea
and the uncertainties on $A$ and $B$ can be obtained by using the symmetric formula
\ba
\Delta{A} = \frac{1}{2}\sqrt{\sum_{k=1}^{n} \left[A(f_k^+) - A(f_k^-)\right]^2}\,.
\ea
The best-fit estimate for $A_0$ is defined as $A(f_0)$ and $f_k^\pm$ represent the $n$ PDF eigenvector sets
in the positive and negative direction respectively.
When $A$ and $B$ are strongly correlated, then $\cos{\phi}_{AB} \approx 1$. Anticorrelation corresponds to
$\cos{\phi}_{AB} \approx -1$, and uncorrelation to $\cos{\phi}_{AB} \approx 0$.
The simultaneous uncertainty boundaries on $A$ and $B$, representing the allowed regions,
can be obtained with the Lissajous parametric ellipse, defined as
\ba
&&A= A_0 + \Delta{A}\sin{\left(\theta+\phi_{AB}\right)}
\nonumber\\
&&B= B_0 + \Delta{B}\sin{\theta}
\ea
where the parameter $\theta$ is in the interval $0 < \theta < 2 \pi$ (see Ref.\cite{Pumplin:2001ct}).

\clearpage

\mysection{\bf Appendix: Additional tables for total cross sections  \label{AppC}}

We provide two tables with aNNLO cross sections with their scale and
PDF uncertainties as well as the associated $K$-factors at 13 TeV LHC energy.
Results are given for three choices of $Z'$ mass.

\begin{table}[!htb]
\begin{center}
\begin{tabular}{|c|c|c|c|c|} \hline
$m_{Z'}$ (TeV)& $\sigma_{aNNLO}$ (fb) &  $\delta$PDF(CT14NNLO) & $\delta$scale & $K$-factor
\\
\hline\hline
1  & 14.4 &  $\pm 0.3$  & $^{+0.3}_{-0.4 }$ &  1.74
\\  \hline
3 &  0.272  &  $^{+0.025 }_{-0.062 }$ & $^{+0.001}_{-0.006}$  &  2.24
\\  \hline
5 & 0.00659 &  $^{+0.00134}_{-0.00086}$  & $^{+0.00001}_{-0.00018}$  & 2.78
\\ \hline
\end{tabular}
\caption[]{aNNLO cross sections and aNNLO/LO $K$-factors for $gu \rightarrow tZ'$
with $k_{tuZ'}/\Lambda=0.01/m_t$ and $m_t=172.5$ GeV at 13 TeV LHC collider energy.
The CT14NNLO PDF uncertainties are calculated at the 68\% C.L.
The scale uncertainties are obtained by taking up and down
variations of the factorization scale $\mu$,  $m_{Z'}/2 <\mu< 2m_{Z'}$.}
\label{tZ'table-gu}
\end{center}
\end{table}

Table 2 shows the aNNLO cross sections for the FCNC process $gu \rightarrow tZ'$ with $k_{tuZ'}/\Lambda=0.01/m_t$.
As shown in Sects. 2.4 and 3, the cross sections are proportional to $k_{tuZ'}^2/\Lambda^2$ so it is trivial
to recalculate them for any other value of $k_{tuZ'}/\Lambda$.

\begin{table}[htb]
\begin{center}
\begin{tabular}{|c|c|c|c|c|} \hline
$m_{Z'}$ (TeV)& $\sigma_{aNNLO}$ (fb) &  $\delta$PDF(NNPDF3.1) & $\delta$scale & $K$-factor
\\
\hline\hline
1 & 157    & $\pm 16  $   & $^{+56 }_{-60 }$    & 2.12
\\  \hline
3 & 0.122    &  $\pm 0.018  $ &  $^{+0.021 }_{-0.026 }$   &  2.66
\\  \hline
5 & 3.34 $\times 10^{-4}$   & $\pm 1.88 \times 10^{-4}$  & $^{+3.7\times 10^{-5}}_{-5.3\times 10^{-5}}$    &  3.33
\\ \hline
\end{tabular}
\caption[]{aNNLO cross sections and aNNLO/LO $K$-factors for $gt \rightarrow tZ'$ with $g_{Z'}=1$ and $m_t=172.5$ GeV at 13 TeV LHC collider energy.
The NNPDF3.1 $n_f = 6$ PDF uncertainties are determined at the 1-$\sigma$ C.L.
The scale uncertainties are obtained by taking up and down variations of the factorization scale $\mu$,  $m_{Z'}/2 <\mu< 2m_{Z'}$.}
\label{tZ'table-gt}
\end{center}
\end{table}

Table 3 shows the aNNLO cross sections for the process $gt \rightarrow tZ'$ with $g_{Z'}=1$.
The dependence of the cross sections on $g_{Z'}$ is given through the formulas in Sects. 2.4 and 3.

\clearpage

\providecommand{\href}[2]{#2}\begingroup\raggedright
\endgroup

\end{document}